\documentclass[twocolumn,superscriptaddress,nofootinbib]{revtex4-2}
\usepackage{hyperref}
\usepackage{bm}
\usepackage{graphicx}
\usepackage{braket}
\usepackage{aas_macros}
\usepackage{amsmath}
\usepackage{amssymb}
\usepackage{ascmac}

\begin{document}
\title{Imprints of primordial magnetic fields on intrinsic alignments of galaxies}
\author{Shohei Saga}
\email{shohei.saga@yukawa.kyoto-u.ac.jp}
\affiliation{Institute for Advanced Research, Nagoya University, Furo-cho Chikusa-ku, Nagoya 464-8601, Japan}
\affiliation{Kobayashi-Maskawa Institute for the Origin of Particles and the
Universe, Nagoya University, Chikusa-ku, Nagoya, 464-8602, Japan}
\affiliation{Sorbonne Universit\'e, CNRS, UMR7095, Institut d'Astrophysique de Paris, 98bis boulevard Arago, F-75014 Paris, France}
\author{Maresuke Shiraishi}
\affiliation{School of General and Management Studies, Suwa University of Science, Chino, Nagano 391-0292, Japan}
\author{Kazuyuki Akitsu}
\affiliation{Theory Center, Institute of Particle and Nuclear Studies, KEK, Tsukuba, Ibaraki 305-0801, Japan}
\affiliation{School of Natural Sciences, Institute for Advanced Study, 1 Einstein Drive, Princeton, NJ 08540, USA}
\author{Teppei Okumura}
\affiliation{Academia Sinica Institute of Astronomy and Astrophysics (ASIAA), No. 1, Section 4, Roosevelt Road, Taipei 10617, Taiwan}
\affiliation{Kavli Institute for the Physics and Mathematics of the Universe (WPI), UTIAS, The University of Tokyo, Kashiwa, Chiba 277-8583, Japan}
\date{\today}
\begin{abstract}
Primordial magnetic fields (PMFs) are one of the plausible candidates for the origin of the observed large-scale magnetic fields. While many proposals have been made for the generation mechanism of PMFs by earlier studies, it remains a subject of debate.
In this paper, to obtain new insights into PMFs, we focus on the intrinsic alignments (IAs) of galaxies induced by the vector and tensor modes of the anisotropic stress of PMFs. 
The long-wavelength vector and tensor modes locally induce the tidal gravitational fields, leading to the characteristic distortions of the intrinsic ellipticity of galaxies.
We investigate the shear E- and B-mode power spectra induced by the magnetic vector and tensor modes in the three-dimensional space, assuming the combination of galaxy imaging and galaxy redshift surveys.
We find that the magnetic tensor mode dominates both the E- and B-mode spectra. In particular, the B-mode spectrum induced by the magnetic tensor mode plays a crucial role in constraining the amplitude of PMFs, even in the presence of the non-magnetic scalar contribution to the B-mode spectrum arising from the one-loop effect.
In future galaxy redshift surveys, such as Euclid and Square Kilometre Array, the minimum detectable value reaches $\sim 30 \, \rm nG$, which can potentially get even smaller in proportion to the number of observed galaxies and reach $\sim \mathcal{O}(1 \, {\rm nG})$. Measuring the IAs of galaxies would be a potential probe for PMFs in future galaxy surveys.
\end{abstract}
\maketitle

\section{Introduction}

Recent observations of high-energy TeV photons emitted from the distant blazars suggest the existence of large-scale magnetic fields, especially in intergalactic and void regions~\cite{2010Sci...328...73N,2011MNRAS.414.3566T,2012ApJ...747L..14V,2013ApJ...771L..42T,2015RAA....15.2173Y,2017ApJ...847...39V}. For instance, Ref.~\cite{2010Sci...328...73N} has reported the lower bound $3\times 10^{-16}$ Gauss on the amplitude of intergalactic magnetic fields. Although the origin of such magnetic fields remains an open question, a number of theories have been proposed to explain them. 
An interesting scenario attracting much attention is the primordial origin, in which the primordial magnetic fields (PMFs) are generated in the early universe, especially before the cosmic recombination epoch. Since PMFs are generated before the formation of stars or galaxies, we expect to observe PMFs as large-scale magnetic fields, not associated with astronomical objects~(see e.g. \cite{2001PhR...348..163G,2013A&ARv..21...62D,2016RPPh...79g6901S} for reviews).

There exists a variety of models for the generation of PMFs. 
In the presence of an interaction between electromagnetic fields and other fields that breaks the conformal invariance during inflation, the inflationary magnetogenesis takes place from the quantum fluctuations~\cite{PhysRevD.37.2743,1992ApJ...391L...1R,1993PhRvD..48.2499D,2004PhRvD..69d3507B,2009JCAP...08..025D,2011PhR...505....1K,2014JCAP...10..056C,2016EL....11519001D,2018CQGra..35j5015B,2019arXiv191110424S,2019JCAP...10..008S,2020PhRvD.102j3508T,2022PhRvD.105b3528T}. The coherent length of PMFs generated in this way can be beyond the horizon scales. During cosmological phase transitions, the bubble collisions and turbulence in the primordial plasma result in generation of PMFs~\cite{VACHASPATI1991258,1997PhRvD..55.4582S,2012ApJ...759...54T}. In the simple phase transition models, the coherent length of PMFs is generally shorter than the observed intergalactic magnetic field. However, recent works~\cite{2019PhRvD.100h3006Z,2019JCAP...09..019E,2021PhRvL.126y1102D,2022PhRvD.106b3510Y} have proposed the model that can generate PMFs with sufficiently long coherent length.
In post-inflationary epochs, the Harrison mechanism~\cite{10.1093/mnras/147.3.279} results in PMFs at $O({\rm Mpc})$ scales in the primordial plasma. However, the amplitude is about $10^{-24}$ Gauss~\cite{2005PhRvL..95l1301T,2011MNRAS.414.2354F,2015PhRvD..91l3510S,2016PhRvD..93j3536F}, which is smaller than the observed amplitude.

To distinguish magnetogenesis models through observations, many authors have investigated the impact of PMFs on cosmological observables, for instance, the effects of PMFs on the Big Bang Nucleosynthesis~\cite{2001PhRvD..65b3517C,2012PhRvD..86l3006Y,2012PhRvD..86f3003K,2019ApJ...872..172L,2021PhRvD.104l3534K,2022ApJ...926L...4L}, cosmic microwave background (CMB) anisotropies~\cite{2000PhRvD..61d3001D,2004PhRvD..70d3011L,2004PhRvD..70l3507G,2008PhRvD..78b3510F,2009MNRAS.396..523P,2010PhRvD..81d3517S,2013PhRvD..88h3515B,2019MNRAS.490.4419S,2019PhRvL.123b1301J,2021JCAP...03..093M,2022PhRvD.105f3537M}, CMB spectral distortions~\cite{2000PhRvL..85..700J,2014JCAP...01..009K,2018MNRAS.474L..52S}, and large-scale structure of the Universe~\cite{2012SSRv..166....1R,2012PhRvD..86d3510S,2012JCAP...11..055F,2014JCAP...03..027C}.
While cosmological observations to date have provided some clues to the generation mechanism of PMFs, it is also worth exploring other ways to extract further information from future cosmological observations.

Motivated by the above, this paper focuses on the intrinsic alignments (IAs) of galaxy shapes as a novel probe for PMFs.
In weak gravitational lensing observations, the IAs of galaxies have been recognized as a contaminant to the estimation of cosmological parameters~\cite{2000MNRAS.319..649H,2000ApJ...545..561C,2001MNRAS.323..713C,2001ApJ...559..552C,2004PhRvD..70f3526H,2006MNRAS.367..611M,2007MNRAS.381.1197H,2009ApJ...694..214O} (also see Ref.~\cite{2015PhR...558....1T} for a review). 
However, it has been shown that the IAs of galaxies offer a unique opportunity to constrain the cosmological parameter, growth of the large-scale structure and the initial condition of the Universe, complementary to galaxy clustering~\cite{2012PhRvD..86h3513S,2013JCAP...12..029C,2019PhRvD.100j3507O,2020MNRAS.493L.124O,2020ApJ...891L..42T,2021MNRAS.503L...6S,2021PhRvD.103h3508A, 2021JCAP...03..060K, 2022PhRvD.106d3523O,2022MNRAS.515.4464C,2023MNRAS.518.4976S,2023JCAP...08..013S,2023PhRvD.108h3533K}. More interestingly, using the galaxy samples in the Sloan Digital Sky Survey, Ref.~\cite{2023ApJ...945L..30O} has measured the anisotropic signals of the IA due to redshift-space distortions and indeed used the signals to constrain the growth rate of the Universe.

How do PMFs leave their imprints on the IA? Refs.~\cite{2010PhRvL.105p1302M,2012PhRvL.108y1301J,2013PhRvD..88d3507D,2014PhRvD..89h3507S,2023PhRvD.107f3531A} have shown that the long-wavelength vector and tensor metric perturbations induce the short-wavelength gravitational tidal fields, called the fossil effect. Such tidal fields are expected to affect the intrinsic shape of galaxies, leading to the IAs.
The imprints of primordial gravitational waves (GWs) on the IAs have been analytically and numerically investigated~\cite{2010PhRvL.105p1302M,2012PhRvL.108y1301J,2013PhRvD..88d3507D,2014PhRvD..89h3507S,2023PhRvD.107f3531A}. Very recently, assuming the photometric surveys, Ref.~\cite{2023arXiv230908653P} has investigated the impact of the primordial vorticity vector mode~\cite{2004PhRvD..70d3518L,2004PhRvD..70d3011L} and primordial GWs on the IAs.
Since our primary interest lies in the IAs as a new probe of PMFs, this paper focuses on the vector and tensor modes induced by PMFs.

The anisotropic stress fluctuation of PMFs creates additional metric perturbations on standard non-magnetic contributions~\cite{2004PhRvD..70d3011L,2008PhRvD..78b3510F,2009MNRAS.396..523P,2010PhRvD..81d3517S}. The resultant metric perturbations source the IAs. As for the scalar mode, the magnetic contribution is not remarkable and hidden by the non-magnetic one; thus, we shall not investigate it below.
By elucidating the relation between the anisotropic stress of PMFs and the IAs, we derive the analytical expression of the E- and B-mode power spectra of the intrinsic ellipticity of galaxies induced by the vector and tensor modes of PMFs. 
Exploiting the derived analytical expression, we explore the potential to constrain the amplitude of PMFs in the Euclid and Square Kilometre Array (SKA) galaxy redshift survey as well as more idealistic cases.

This paper is organized as follows. In Sec.~\ref{sec: intro vector and tensor}, we briefly review the general properties of PMFs and present how the anisotropic stress of PMFs induces the vector and tensor modes. For comparison purposes, we also introduce the vorticity vector mode and primordial GWs. In Sec.~\ref{sec: fossil}, based on Ref.~\cite{2014PhRvD..89h3507S}, we show the analytical expression of the intrinsic ellipticity shape induced by the long-wavelength vector and tensor modes. We show the detailed derivations in Appendix~\ref{app: fossil}.
In Sec.~\ref{sec: E and B modes}, we derive an analytical expression for the three-dimensional E- and B-mode power spectra, and see their typical behavior. Using the analytical expression of the E- and B-mode power spectra, we perform the Fisher matrix computation and derive the expected minimum detectable value of the amplitude of PMFs in future surveys in Sec.~\ref{sec: Fisher}.
In our analysis, we take into account the non-magnetic scalar mode contributions to the E- and B-mode power spectra, respectively arising from the leading-order and one-loop order effects. We present the way to compute the one-loop B-mode spectrum in Appendix~\ref{app: 1loop}.
We perform the same analysis for the vorticity vector mode and primordial GWs in Appendix~\ref{app: others}. Sec.~\ref{sec: summary} is devoted to the summary of our findings. 
Throughout this paper, we apply the Einstein summation convention for repeated Greek indices and alphabets running from 0 to 3 and from 1 to 3, respectively. We work in units $c=\hbar=1$.

\section{Vector and tensor modes\label{sec: intro vector and tensor}}

We are interested in imprints of the vector and tensor modes induced by PMFs on cosmological observables.
In this section, we briefly introduce the basic property of the initial power spectrum of PMFs and the vector and tensor modes induced by PMFs in Sec.~\ref{sec: PMFs}. We also introduce other possible cosmological sources to generate the vector and tensor modes in Sec.~\ref{sec: other VT} for comparison purposes.

Throughout this paper, we work in the synchronous gauge of which the line element is
\begin{align}
{\rm d}s^{2} = a^{2}(\eta) \left[ - {\rm d}\eta^{2} + \left( \delta_{ij} + h_{ij} \right) {\rm d}x^{i}{\rm d}x^{j} \right] , \label{eq: metric}
\end{align}
where the quantities $a$ and $\eta$ are the scale factor and the conformal time, respectively. We decompose the metric perturbation into the vector and tensor modes based on the helicity basis in Fourier space:
\begin{align}
h_{ij} = 
\sum_{\lambda=\pm1}O^{(\lambda)}_{ij}h^{(\lambda)}(\bm{k})
+ \sum_{\lambda=\pm2}O^{(\lambda)}_{ij}h^{(\lambda)}(\bm{k}) , \label{eq: svt decomposition}
\end{align}
where the first and second terms represent the vector ($\lambda=\pm1$) and tensor ($\lambda=\pm2$) modes, respectively.
Here we have defined
\begin{align}
O^{(\pm1)}_{ij}(\hat{\bm{k}}) & = \hat{k}_{i}\epsilon^{(\pm1)}_{j}(\hat{\bm{k}}) + \hat{k}_{j}\epsilon^{(\pm1)}_{i}(\hat{\bm{k}}) , \\
O^{(\pm2)}_{ij}(\hat{\bm{k}}) & = \epsilon^{(\pm1)}_{i}(\hat{\bm{k}})\epsilon^{(\pm1)}_{j}(\hat{\bm{k}}) ,
\end{align}
where the polarization vectors $\bm{\epsilon}^{(\pm1)}$ satisfy the relations: $\hat{\bm{k}}\cdot\bm{\epsilon}^{(\pm1)} = 0$, $\left(\bm{\epsilon}^{(\pm1)}\right)^{*} = \bm{\epsilon}^{(\mp1)}$, and $\bm{\epsilon}^{(\pm1)} \cdot \bm{\epsilon}^{(\mp1)} = 1$.
For the vector mode, we define the gauge invariant variable in Fourier space by 
\begin{align}
\sigma_{ij}(\bm{k}) \equiv \sum_{\lambda=\pm1}O^{(\lambda)}_{ij}h^{(\lambda)\prime}_{ij}(\bm{k})/k
,
\end{align}
where a prime denotes a derivative with respect to the conformal time $\eta$. With this definition, the helicity modes of $\sigma_{ij}$ are given by $\sigma^{(\pm1)} = {h^{(\pm1)}}'/k$.

\subsection{Magnetic vector and tensor modes
\label{sec: PMFs}}

PMFs act as a source of the vector and tensor modes~\cite{2002PhRvD..65l3004M,2004PhRvD..70d3011L,2008PhRvD..77f3003G,2008PhRvD..78b3510F,2009MNRAS.396..523P,2010PhRvD..81d3517S}.
We first introduce the general property of PMFs. We then present the magnetic vector and tensor modes induced by PMFs.

\subsubsection{General property of PMFs\label{eq: general PMFs}}

We consider the magnetically induced vector and tensor modes presented by Refs.~\cite{2004PhRvD..70d3011L,2008PhRvD..78b3510F,2009MNRAS.396..523P,2010PhRvD..81d3517S}.
We assume that the time evolution of physical magnetic fields $B(a,\bm{x})$ is given by $B(a,\bm{x}) = B(\bm{x})/a^{2}$ with $B(\bm{x})$ being the comoving magnetic fields without the adiabatic decay due to the expansion of the Universe. This assumption is valid in the limit of infinite electrical conductivity of the Universe as in the early time.

The power spectrum of the divergence-free vector such as PMFs is given by
\begin{align}
\Braket{B_{i}(\bm{k})B^{*}_{j}(\bm{k}')} = (2\pi)^{3}\delta^{3}_{\rm D}(\bm{k}-\bm{k}')P_{ij}(\hat{\bm{k}})P_{B}(k) ,
\end{align}
where $P_{ij}(\hat{\bm{k}}) = (\delta_{ij}- \hat{k}_{i}\hat{k}_{j})/2$ and $B_{i}(\bm{k})$ is the Fourier transform of $B_{i}(\bm{x})$ given by
\begin{align}
\bm{B}(\bm{x}) = \int\frac{{\rm d}^{3}\bm{k}}{(2\pi)^{3}}\bm{B}(\bm{k}) e^{i\bm{k}\cdot\bm{x}} .
\end{align}
We model the power spectrum of the primordial magnetic field by the power-law form (see e.g. Ref.~\cite{2016A&A...594A..19P}):
\begin{align}
P_{B}(k) = B^{2}_{\lambda}\frac{\Gamma\left( \frac{n_{B}+3}{2} \right)}{4\pi^{2}\lambda^{n_{B}+3}}\, k^{n_{B}} \Theta(k_{\rm D}-k) , \label{eq: PMF power spectrum}
\end{align}
where the functions $\Gamma(x)$ and $\Theta(x)$ are the Gamma function and the Heaviside theta function, respectively. The quantities $B_{\lambda}$ and $k_{\rm D}$ are the amplitude of PMFs smoothed over a comoving scale of $\lambda = 1\, {\rm Mpc}$ and the damping scale, respectively. We introduce the Heaviside theta function to express the damping scale. We use the damping scale $k_{\rm D}$ given by Ref.~\cite{1998PhRvD..58h3502S,2002PhRvD..65l3004M,2016A&A...594A..19P}:
\begin{align}
k_{\rm D} = (2.9\times 10^{4})^{\frac{1}{n_{B}+5}}\left( \frac{B_{\lambda}}{\rm nG}\right)^{-\frac{2}{n_{B}+5}} (2\pi)^{\frac{n_{B}+3}{n_{B}+5}}h^{\frac{1}{n_{B}+5}}, \label{eq: kD}
\end{align}
with $h$ being the reduced Hubble constant.
The power spectrum of PMFs contains two parameters, $B_{\lambda}$ and $n_{B}$. The CMB observations provide the upper limits $B_{\lambda} \lesssim O(1\, {\rm nG})$ depending on the value of $n_{B}$.

A key quantity for investigating the cosmological impact of PMFs is the anisotropic stress and its power spectrum.
The anisotropic stress of PMFs, $\Pi_{B, ij}$, is given by
\begin{align}
\Pi_{B, ij}(\bm{k}) &= -\frac{1}{4\pi\rho_{\gamma,0}}\int\frac{{\rm d}^3 \bm{k}_{1}}{(2\pi)^{3}}\, B_{i}(\bm{k}_{1})B_{j}(\bm{k}-\bm{k}_{1}) , \label{eq: Pi ij}
\end{align}
where $\rho_{\gamma0}$ is the photon energy density at the present time.
As with the metric perturbation, we decompose the anisotropic stress into the vector and tensor components by using the helicity basis:
\begin{align}
\Pi_{B,ij}(\bm{k}) = 
\sum_{\lambda=\pm1}O^{(\lambda)}_{ij}(\hat{\bm{k}}) \Pi^{(\lambda)}_{B} (\bm{k}) + 
\sum_{\lambda=\pm2}O^{(\lambda)}_{ij}(\hat{\bm{k}}) \Pi^{(\lambda)}_{B} (\bm{k})\label{eq: Pi 1} .
\end{align}
The power spectra of the anisotropic stress for the vector and tensor modes are then given by~\cite{2004PhRvD..70d3011L,2010PhRvD..81d3517S}
\begin{align}
\Braket{\Pi^{(\lambda)}_{B} (\bm{k}) \Pi^{(-\lambda)}_{B} (\bm{k}')}&= (2\pi)^{3}\delta^{3}_{\rm D}(\bm{k}-\bm{k}') \frac{1}{2}|\Pi^{(X)}_{B}|^{2}, \label{eq: def Pi}
\end{align}
where we denote $X=V$ and $X=T$ for the vector ($\lambda = \pm1$) and tensor ($\pm2$) modes, respectively. Here we assume the unpolarized case, where the power spectra of the $+$ and $-$ modes are identical. In Eq.~(\ref{eq: def Pi}), we define
\begin{align}
|\Pi^{(V/T)}_{B}(\bm{k})|^{2} &= 
\frac{c_{V/T}}{4(4\pi\rho_{\gamma0})^{2}} \int\frac{{\rm d}^{3}\bm{k}_{1}}{(2\pi)^{3}}
\notag \\
& \qquad  \times 
P_{B}(k_{1})P_{B}(k_{2}) D_{V/T}(k,k_{1},\mu) , \label{eq: P Pi}
\end{align}
where $\bm{k}_{2} = \bm{k} - \bm{k}_{1}$. 
In the above, $c_{V/T}$ are constants, $c_{V} = 1$ and $c_{T} = 1/2$, and we have defined $D_{V}(k,k_{1},\mu) = 1-2\gamma^{2}\beta^{2}+\gamma\beta\mu$, and $D_{T}(k,k_{1},\mu) = (1 + \gamma^{2})(1+\beta^{2})$, with $\mu = \hat{\bm{k}}_{1}\cdot\hat{\bm{k}}_{2}$, $\gamma = \hat{\bm{k}}\cdot\hat{\bm{k}}_{1}$,  $\beta = \hat{\bm{k}}\cdot\hat{\bm{k}}_{2}$.

The anisotropic stress defined in Eq.~(\ref{eq: Pi ij}) contributes to the energy-momentum tensor in the Einstein equation, and sources the fluctuations induced magnetically. In what follows, we briefly review the resultant vector and tensor fluctuations induced by the anisotropic stress of PMFs.

\subsubsection{Vector mode\label{sec: compensated}}

We first introduce the magnetically induced vector mode. In the initial time, the anisotropic stress of PMFs is compensated by that of neutrinos. Hence, the total anisotropic stress is zero, but each component is perturbed. In this setup, Ref.~\cite{2010PhRvD..81d3517S} has derived the initial condition for the Einstein-Boltzmann equation and shown that all the perturbed variables are proportional to $\Pi_{B}^{(\pm 1)}(\bm{k})$ in Eq.~(\ref{eq: Pi 1}) (see Appendix~B in Ref.~\cite{2010PhRvD..81d3517S}). This contribution is known as \textit{compensated vector mode}. Here, we define the transfer function of the vector mode by
\begin{align}
\sigma^{(\pm1)} (\eta, \bm{k}) & = \mathcal{T}^{(V)}_{B}(\eta, k) \Pi^{(\pm1)}_{B}(\bm{k}). \label{eq: sigma pm1 mag}
\end{align}
We note that even though the magnetic field itself is a random Gaussian variable, the statistical property of the vector mode is highly non-Gaussian, as the anisotropic stress is proportional to the square of the magnetic field.

\begin{figure*}
\centering
\includegraphics[width=0.9\textwidth]{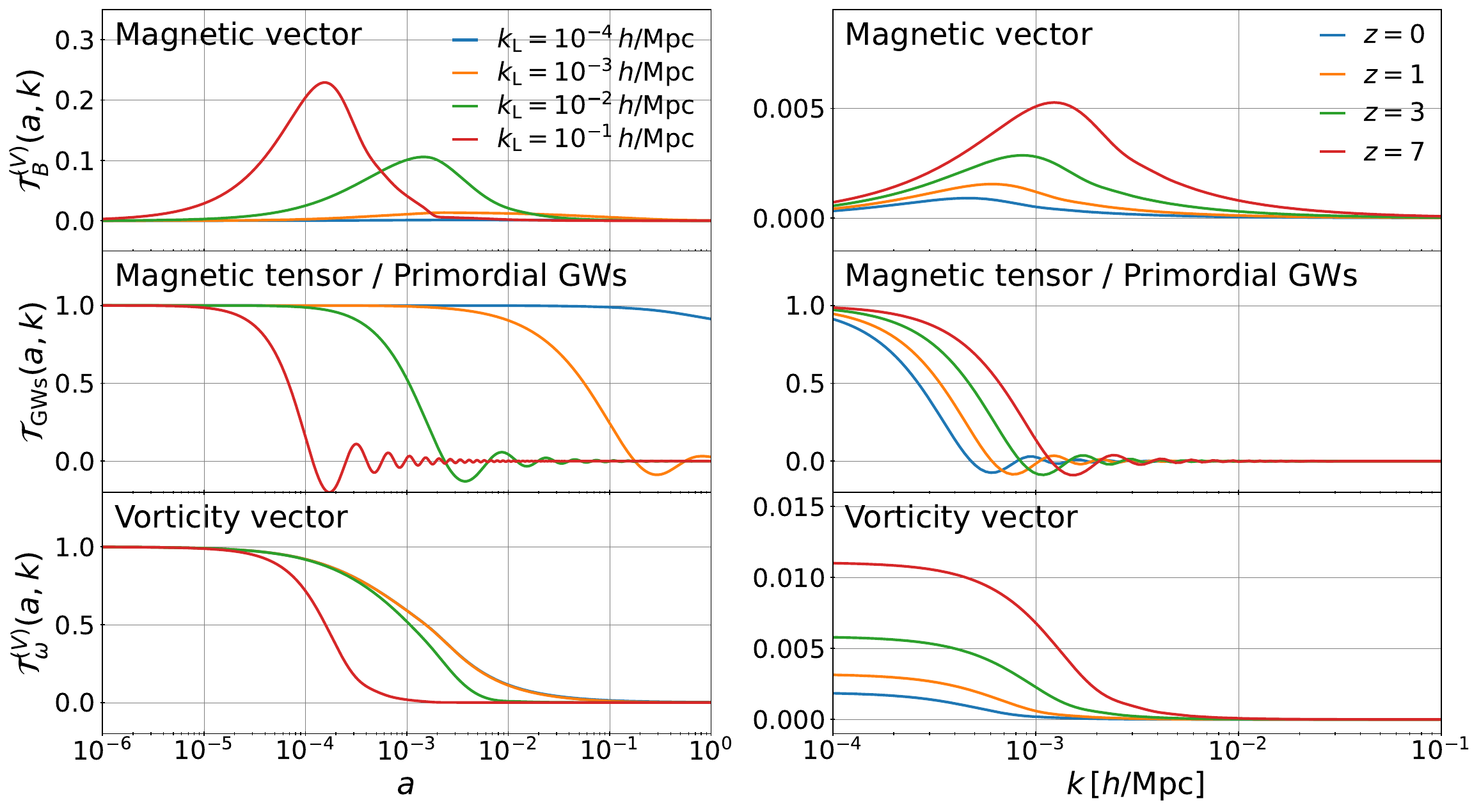}
\caption{Time and wavenumber dependences (left and right, respectively) of the transfer function. From top to bottom, we show the transfer function of the magnetic vector mode~\cite{2002PhRvD..65l3004M,2004PhRvD..70d3011L}, magnetic tensor mode/primordial GWs, and vorticity vector mode~\cite{2010PhRvD..81d3517S}, respectively.}
\label{fig: transfer}
\end{figure*}

We plot the time and wavenumber dependences of the transfer function in the top panels of Fig.~\ref{fig: transfer}. To compute the transfer function of the compensated vector mode $\mathcal{T}^{(V)}_{B}$, we use the {\tt CAMB} code~\cite{2000ApJ...538..473L}.\footnote{In the initial parameter file for the {\tt CAMB} code, we set \texttt{vector\_mode$=$1} to output the transfer function of the compensated vector mode.} In the right panel, we show only the low-redshift results because we are interested in the late-time effect of PMFs on the galaxy IAs.
In the limit $a\to 0$, in contrast to the usual adiabatic initial condition or primordial GWs cases, the transfer function asymptotically approaches zero. This is because the vector metric perturbation is sourced by the total anisotropic stress, which however initially cancels between the magnetic field and neutrinos.

\subsubsection{Tensor mode\label{sec: passive}}

The tensor metric perturbation arises from the anisotropic stress of PMFs after its generation time. However, after the neutrino decoupling, the contribution to the total energy-momentum tensor from the anisotropic stress of PMFs is cancelled by the anisotropic stress of neutrinos. The generation of the tensor mode therefore ceases at the epoch of neutrino decoupling. This contribution is known as \textit{passive tensor mode} \cite{2010PhRvD..81d3517S}.

The expression of generated tensor mode is given by~\cite{2010PhRvD..81d3517S}:
\begin{align}
h^{(\pm 2)}_{\rm ini}(\bm{k}) & =  6 R_{\gamma}\ln{\left(\eta_{\nu}/\eta_{B}\right)} \Pi^{(\pm2)}_{B}(\bm{k}), \label{eq: passive}
\end{align}
where we define an energy fraction of photons in the total radiation defined $R_{\gamma} = \rho_{\gamma}/(\rho_{\gamma}+\rho_{\nu})$, and the quantities $\eta_{\nu}$ and $\eta_{B}$ are the neutrino decoupling time and the PMF generation time in terms of the conformal time, respectively. 
The generation epoch $\eta_{B}$ highly depends on the generation mechanism; therefore, $\eta_{\nu}/\eta_{B}$ has an ambiguity as $10^{6}\lesssim \eta_{\nu}/\eta_{B}\lesssim 10^{17}$. In the following analysis, we adopt the maximum value $\eta_{\nu}/\eta_{B} = 10^{17}$, corresponding to the grand unification energy scale~\cite{2016A&A...594A..19P}, while the change of $h^{(\pm 2)}_{\rm ini}$ (and induced E/B-mode ellipticity field appearing below) is only by a factor of $< 3$ even if adopting other value.

Once the tensor metric perturbation is generated from the anisotropic stress of PMFs, it evolves as the usual primordial GWs~\cite{2010PhRvD..81d3517S,2000PhRvD..61d3001D,2018PhRvD..98h3518S}. Therefore we may express the magnetic tensor mode at the time $\eta$ by using the transfer function of the primordial GWs $\mathcal{T}_{\rm GWs}(k,\eta)$ by
\begin{align}
h^{(\pm 2)}(\eta, \bm{k}) & =  6 R_{\gamma}\ln{\left(\eta_{\nu}/\eta_{B}\right)} \mathcal{T}_{\rm GWs}(k,\eta) \Pi^{(\pm2)}_{B}(\bm{k}) , \label{eq: passive 2}
\end{align}
where $\mathcal{T}_{\rm GWs}(k,\eta)$ stands for the transfer function of the primordial GWs as explained in the next subsection.
As with the magnetic vector mode, the magnetic tensor mode is proportional to the square of PMFs, suggesting that its statistical property is highly non-Gaussian.
We remark that the tensor mode arises from the same mechanism in Sec.~\ref{sec: compensated}, i.e., \textit{compensated tensor mode}. However, its amplitude are negligible compared to the tensor mode presented in this section~(see e.g. Ref.~\cite{2016A&A...594A..19P}), and hence we ignore it throughout this paper.

\subsection{Other vector and tensor sources
\label{sec: other VT}}

Here, we introduce other possible sources of vector and tensor modes: the (neutrino) vorticity vector mode and the primordial GWs. In the next section, we will compare the vector and tensor modes introduced in this subsection with those induced by PMFs.

Another possible source of a vector mode is the (neutrino) vorticity vector mode~\cite{2004PhRvD..70d3518L,2004PhRvD..70d3011L,2010PhRvD..81d3517S}, in which, similar to the isocurvature initial conditions, the sum of the neutrino, baryon, and photon vorticities vanishes initially, but the vector metric perturbation remains constant due to the neutrino anisotropic stress (see Appendix~2 in Ref.~\cite{2010PhRvD..81d3517S}). 
Using the transfer function of the vorticity mode $\mathcal{T}^{(V)}_{\omega}(\eta, k)$, the vector metric perturbation at the time $\eta$ is given by
\begin{align}
\sigma^{(\pm1)} (\eta, \bm{k}) & = \mathcal{T}^{(V)}_{\omega}(\eta, k) \sigma^{(\pm1)}_{\rm ini} (\bm{k}) ,
\end{align}
where $\sigma^{(\pm1)}_{\rm ini} (\bm{k})$ being the primordial amplitude. We define its power spectrum by
\begin{align}
\Braket{\sigma^{(\pm1)}_{\rm ini} (\bm{k}) \sigma^{(\mp1)}_{\rm ini} (\bm{k}')} = (2\pi)^{3}\delta^{3}_{\rm D}(\bm{k}-\bm{k}')\frac{2\pi^{2}}{k^{3}}\mathcal{P}_{\sigma^{(\pm1)}}(k) .
\end{align}
In the unpolarized case $\mathcal{P}_{\sigma^{(+1)}}(k) = \mathcal{P}_{\sigma^{(-1)}}(k) = \mathcal{P}_{\sigma}(k)/2$, where $\mathcal{P}_{\sigma}(k)$ stands for total power spectrum defined by $\Braket{\sigma_{{\rm ini}\,i} (\bm{k}) \sigma^{*}_{{\rm ini}\,i} (\bm{k}')} = (2\pi)^{3}\delta^{3}_{\rm D}(\bm{k}-\bm{k}')(2\pi^{2}/k^{3})\, \mathcal{P}_{\sigma}(k)$. We parametrize the total power spectrum by the power-law form:
\begin{align}
\mathcal{P}_{\sigma} (k) & = r_{V}\mathcal{A}_{S}\left( \frac{k}{k_{*}}\right)^{n_{V}},
\end{align}
with $\mathcal{A}_{S}$ and $k_{*} = 0.002\, {\rm Mpc}$ being the amplitude of the usual non-magnetic scalar perturbation and the pivot scale. The shape of the power spectrum is controlled by the vector-to-scalar ratio $r_{V}$ and spectral index $n_{V}$.
Although finding a mechanism to source this mode is challenging, this initial condition is mathematically possible and has been indeed investigated by many authors, e.g., Refs.~\cite{2012PhRvD..85d3009I,2014JCAP...10..004S,2023arXiv230111344C,2023arXiv230908653P}.
We show the behaviors of the transfer function in the middle two panels of Fig.~\ref{fig: transfer} by using the {\tt CAMB} code~\cite{2000ApJ...538..473L}.\footnote{We set \texttt{vector\_mode$=$0} to output the transfer function of the vorticity mode in the initial parameter file.}

Another possible source of a tensor mode is the usual primordial GWs generated during inflation from the quantum fluctuations. The tensor mode is formally given by using the transfer function $\mathcal{T}_{\rm GWs}(\eta, k)$ and initial fluctuation $h^{(\pm2)}_{\rm ini}(\bm{k})$:
\begin{align}
h^{(\pm2)}(\eta, \bm{k}) = \mathcal{T}_{\rm GWs}(\eta, k) h^{(\pm2)}_{\rm ini}(\bm{k}) .
\end{align}
We define the power spectrum of the initial field by
\begin{align}
\Braket{h^{(\pm2)}_{\rm ini} (\bm{k}) h^{(\mp2)}_{\rm ini} (\bm{k}')} = (2\pi)^{3}\delta^{3}_{\rm D}(\bm{k}-\bm{k}')\frac{2\pi^{2}}{k^{3}}\mathcal{P}_{h^{(\pm2)}}(k) .
\end{align}
We consider unpolarized case $\mathcal{P}_{h^{(+2)}}(k) = \mathcal{P}_{h^{(-2)}}(k) = \mathcal{P}_{h}(k)/2$, where $\mathcal{P}_{h}(k)$ stands for total power spectrum defined by $\Braket{h_{{\rm ini}\,ij} (\bm{k}) h^{*}_{{\rm ini}\,ij} (\bm{k}')} = (2\pi)^{3}\delta^{3}_{\rm D}(\bm{k}-\bm{k}')(2\pi^{2})/k^{3}\, \mathcal{P}_{h}(k)$.
We model the total power spectrum of the initial field by the power-law form:
\begin{align}
\mathcal{P}_{h} (k) & = r_{T}\mathcal{A}_{S}\left( \frac{k}{k_{*}}\right)^{n_{T}},
\end{align}
where $r_{T}$ and $n_{T}$ being the usual tensor-to-scalar ratio and spectral index, respectively.
In the bottom panels in Fig.~\ref{fig: transfer}, we show the behaviors of the transfer function $\mathcal{T}_{\rm GWs}$ by numerically solving the evolution equation for the primordial GWs:
\begin{align}
\mathcal{T}''_{\rm GWs}(\eta,k) + 2\mathcal{H} \mathcal{T}'_{\rm GWs}(\eta,k) + k^{2}\mathcal{T}_{\rm GWs}(\eta,k) = 0,
\end{align}
with the initial conditions $\mathcal{T}_{\rm GW}(0,k) = 1$ and $\mathcal{T}'_{\rm GW}(0,k) = 0$. Here, we define the conformal Hubble parameter $\mathcal{H} = a'/a$.

\section{Impact of the vector and tensor modes on the intrinsic alignment\label{sec: fossil}}

\begin{figure*}
\centering
\includegraphics[width=0.9\textwidth]{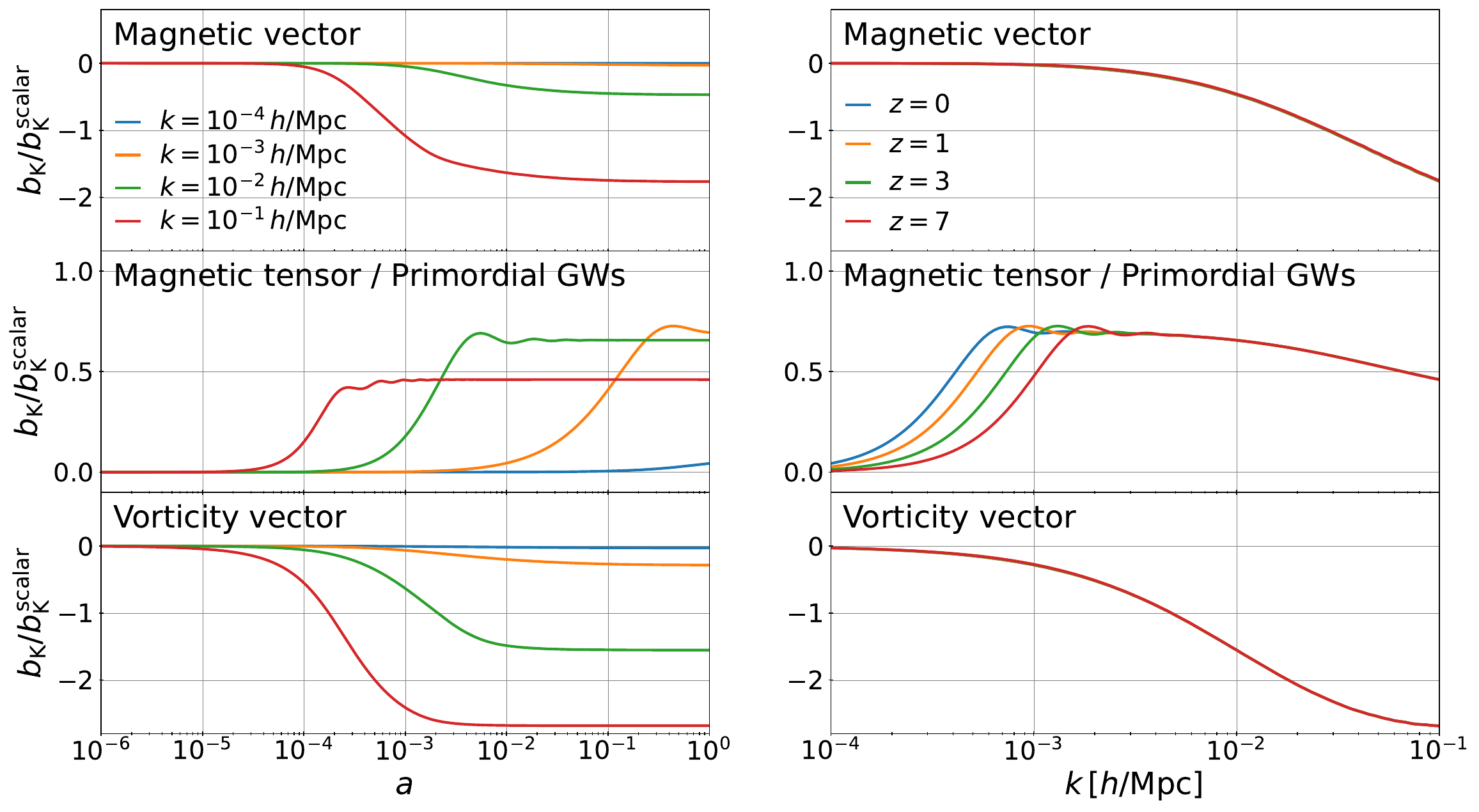}
\caption{Time and wavenumber dependences of the effective bias parameter normalized by the scalar tidal bias $b_{\rm K}/b^{\rm scalar}_{\rm K}$ (left and right, respectively). From top to bottom, we show the magnetic vector mode, magnetic tensor mode/primordial GWs, and vorticity vector mode, respectively.}
\label{fig: bias K}
\end{figure*}

We briefly review how the vector and tensor modes induce the IAs of galaxies.  We leave the detailed derivation to the Appendix~\ref{app: fossil}.

The local physical effects of the long-wavelength vector and tensor modes on the gravitational potential have been investigated by using conformal Fermi normal coordinates in Refs.~\cite{2013PhRvD..88h3502P,2014PhRvD..89h3507S,2015JCAP...11..043D}. 
According to the results shown in Ref.~\cite{2014PhRvD..89h3507S} (or see Appendix~\ref{app: fossil}), the tidal fields locally induced by the vector and tensor modes of the long-wavelength mode $k_{\rm L}$ are, respectively, given by:
\begin{align}
\tau^{(V)}_{ij}(\eta, k_{\rm L}) & = - \frac{k_{\rm L}}{2a}\left( a \sigma_{ij}(\eta, k_{\rm L})  \right)' \notag \\
& =  - \frac{k_{\rm L}}{2a}\left( a \mathcal{T}^{(V)}(\eta,k_{\rm L}) \right)' \sigma_{{\rm ini}\, ij}(k_{\rm L}) 
, \label{eq: tau ij v}\\
\tau^{(T)}_{ij}(\eta, k_{\rm L}) 
& = - \frac{1}{2a}\left( a h'_{ij}(\eta,k_{\rm L})  \right)' \notag \\
& = - \frac{1}{2a}\left( a \mathcal{T}^{(T)\prime}(\eta,k_{\rm L}) \right)' h_{{\rm ini}\,ij}(k_{\rm L}) 
, \label{eq: tau ij t}\end{align}
where we used the gauge invariant vector variable $\sigma_{ij}(k_{\rm L}) = h'_{ij}(k_{\rm L})/k_{\rm L}$.
From the first line to the second line in Eqs.~(\ref{eq: tau ij v}) and (\ref{eq: tau ij t}), we decompose the perturbed metric into the time-dependent part described by the transfer function and the time-independent initial part.

To derive the expression for the density field induced by the coupling between long- and short-wavelength modes, we solve the equation of motion of a matter particle in the local frame:
\begin{align}
\bm{x}'' + \mathcal{H}\bm{x}' &= -\bm{\nabla}_{x}\left( 
\phi_{\rm s} + \frac{1}{2}\tau^{(X)}_{ij}x^{i}x^{j}\right) , \\
\nabla^{2}_{x}\phi_{\rm s} &= 4\pi G a^{2} \bar{\rho}_{\rm m} \delta ,
\end{align}
with $X = V$ and $T$ for the vector and tensor modes, respectively. The quantity $\phi_{\rm s}$ stands for the scalar gravitational potential of the short-wavelength mode.
To facilitate the computations, we employ the Lagrangian perturbation formalism (see Appendix~\ref{app: fossil}).

The second order solution arising from the short- and long-wavelength modes coupling is then given by
\begin{align}
\delta^{(sl)}(\bm{x})
&=
\xi_{{\rm ini}\, ab}(k_{\rm L})
\Biggl[
\left( - \frac{D^{(sl)}(\eta,k_{\rm L}) }{D(\eta)}  + \beta(\eta,k_{\rm L}) \right)
\notag \\
& \qquad 
\times \partial^{-2}\partial_{a}\partial_{b}
+ \beta(\eta,k_{\rm L}) x_{a} \partial_{b}
\Biggr]
\delta^{(1)}(\bm{x},\eta) , \label{eq: 2nd order delta}
\end{align}
where $\xi = \sigma$ or $h$ for the vector or tensor modes, respectively.
We define the linear growth factor $D(\eta)$, which satisfies
\begin{align}
D''(\eta) + \mathcal{H}D'(\eta) - 4\pi G a^{2} \bar{\rho}_{\rm m}(\eta) D(\eta) &= 0  .
\end{align}
The second-order growth factor $D^{(sl)}$ satisfies
\begin{align}
D^{(sl)\prime\prime} + \mathcal{H} D^{(sl)\prime} - 4\pi G a^{2} \bar{\rho}_{\rm m}(\eta) D^{(sl)}
&= S^{(X)}(\eta, k_{\rm L})
\end{align} 
where the source terms $S^{(X)}$ of the vector ($X=V$) and tensor ($X=T$) are, respectively, defined by
\begin{align}
S^{(V)}(\eta, k_{\rm L}) &= -\frac{k_{\rm L}}{2a}D(\eta)(a\mathcal{T}^{(V)}(\eta,k_{\rm L}))' , \label{eq: source v}\\
S^{(T)}(\eta, k_{\rm L}) &= -\frac{1}{2a} D(\eta) (a\mathcal{T}^{(T)\prime}(\eta,k_{\rm L}))' . \label{eq: source t}
\end{align}

Eq.~(\ref{eq: 2nd order delta}) allows us to estimate how the IA is induced by the long-wavelength vector and tensor modes. 
We use the same ansatz in Ref.~\cite{2014PhRvD..89h3507S}, in which we assume that the first term in the square brackets in Eq.~(\ref{eq: 2nd order delta}) induces the intrinsic galaxy shape, and the conversion from the second-order density fluctuations to the intrinsic galaxy shape in the vector and tensor modes has the same scaling as that in the scalar mode~\cite{2014PhRvD..89h3507S,2023PhRvD.107f3531A} (see Appendix~\ref{app: fossil} for details). Finally, we have the expression of the intrinsic galaxy shape induced by the long-wavelength vector and tensor modes at the linear order given by
\begin{align}
\gamma_{ij}(\bm{k}) &= b_{\rm K}(\eta,k) \xi_{{\rm ini}\, ij}(\bm{k}) , \label{eq: gamma ij}
\end{align}
where $\xi = \sigma$ or $h$ for the vector or tensor modes, respectively. Here we define the effective linear shape bias by
\begin{align}
b_{\rm K}(\eta,k) & \equiv \frac{7}{4} \left( - \frac{D^{(sl)}(\eta,k)}{D(\eta)} + \beta(\eta,k) \right) b^{\rm scalar}_{\rm K} .
\end{align}
The quantity $b_{\rm K}^{\rm scalar}$ is the linear shape bias induced by the scalar tides as in Ref.~\cite{2023PhRvD.107f3531A}. We omit the subscript~${}_{\rm L}$ indicating the long-wavelength mode here.

In Fig.~\ref{fig: bias K}, we show the behaviors of the effective linear shape bias parameter $b_{\rm K}(\eta, k)/b^{\rm scalar}_{\rm K}$. 
We notice that the effective linear shape bias sourced by the vector mode has the opposite sign to that by the tensor mode because of the different number of time derivatives in the sources (see Eqs.~(\ref{eq: source v}) and (\ref{eq: source t})).
Taking the limit $k\to \infty$, the transfer function asymptotically approaches zero (see the right panels in Fig.~\ref{fig: transfer}). However the effective tidal bias parameter does not vanish in the same limit, known as the fossil effect~\cite{2014PhRvD..89h3507S}. 
Since the transfer function of the vector modes decays rapidly after the matter-radiation equality, the effective linear shape bias parameter induced by the vector mode, a fossil effect, is more quickly frozen than that by the tensor mode. Therefore, we do not see the redshift dependence of the effective tidal bias parameter in the top and bottom panels in Fig.~\ref{fig: bias K}.
We note that, since the amplitude of the effective linear shape bias parameter is solely determined by the behavior of the transfer function, its amplitude can be larger than the linear shape bias induced by the scalar tides as seen in the vector mode cases. However, the net impact of the vector mode on the galaxy shape is generally much smaller than the scalar one.

While Eq.~(\ref{eq: gamma ij}) is estimated based on the ansatz we mentioned before, it is non-trivial whether Eq.~(\ref{eq: gamma ij}) actually holds in the presence of long-wavelength vector/tensor modes. However, a recent numerical work~\cite{2023PhRvD.107f3531A} has investigated the validity of Eq.~(\ref{eq: gamma ij}) in the primordial GWs case, and confirmed that Eq.~(\ref{eq: gamma ij}) agrees with simulations at large scales. Although they also confirmed that the discrepancy between the ansatz and measurements becomes larger at smaller scales, this discrepancy does not qualitatively affect the estimates in this paper, and henceforth, we will carry out the analysis assuming that Eq.~(\ref{eq: gamma ij}) is valid for both the vector and the tensor modes. Numerical validation, especially for the vector mode, would be an interesting future work.

\section{Three-dimensional E- and B-mode power spectra\label{sec: E and B modes}}

From here, we move to the heart of this work, which involves the investigation of the three-dimensional power spectrum of the IA induced by the vector and tensor modes. While the vector and tensor can generally contribute to the lens-induced ellipticity~(e.g. Refs.~\cite{2003PhRvL..91b1301D,2006PhRvD..74b3521L,2012JCAP...10..030Y,2013JCAP...08..051Y,2015PhRvD..92f3533S}), as we are interested in the characteristic signature of the intrinsic galaxy shapes from the magnetically induced vector and tensor modes, we simply ignore the lens-induced ellipticity but focus on the intrinsic ellipticity.

The E- and B-mode decomposition is convenient to distinguish the impact of vector and tensor modes on the galaxy shape from that of the scalar mode because the leading-order scalar mode does not induce the B mode. There however exists a scalar-mode contribution to the B mode arising from the one-loop effect, which we will properly take into account in later section.

We work under the plane-parallel limit and set the line-of-sight direction to be parallel to the $z$-axis: $\hat{\bm{n}} = \hat{\bm{z}}$. We define the shear E-mode and B-mode by
\begin{align}
{}_{\pm2}\gamma(\bm{k})e^{\mp2 i\phi_{k}} &= E(\bm{k}) \pm i B(\bm{k})  .
\end{align}
where we define
\begin{align}
{}_{\pm2}\gamma(\bm{k}) &= m^{i}_{\mp} m^{j}_{\mp} \gamma_{ij}(\bm{k}). \label{eq: gamma pm 2}
\end{align}
with $\bm{m}_{\lambda} = (1, -\lambda i , 0)/\sqrt{2}$.
Recall the expression of the induced galaxy shape by the long-wavelength vector and tensor modes in Eq.~(\ref{eq: gamma ij})
\begin{align}
\gamma_{ij} &= b_{\rm K}(k) \xi_{{\rm ini}\, ij}(k) ,\notag \\
& = b_{\rm K}(k) \sum_{\lambda} O^{(\lambda)}_{ij}(\hat{\bm{k}}) \xi^{(\lambda)}_{{\rm ini}}(\bm{k}) , \label{eq: projection}
\end{align}
where $\xi=\sigma$ and $h$ for the vector and tensor modes, respectively. Here and hereafter we 
omit the time dependence of $b_{\rm K}$ to simplify the notation.
Using this expression, we obtain
\begin{align}
{}_{\pm2}\gamma^{(V)} &= 
\frac{b_{\rm K}(k)}{\sqrt{2}}\sum_{\lambda=\pm1}\sigma^{(\lambda)}
\sin{\theta_{k}}(\pm \lambda + \cos{\theta_{k}})e^{\pm 2i\phi_{k}}
\label{eq: gamma vector} , \\
{}_{\pm2}\gamma^{(T)} &= \frac{b_{\rm K}(k)}{4} \sum_{\lambda=\pm2}h^{(\lambda)} \left[ (1+\cos^{2}{\theta_{k}}) \pm 2\lambda \cos\theta_{k} \right] e^{\pm2 i\phi_{k}}
\label{eq: gamma tensor}  ,
\end{align}
where the superscripts $(V)$ and $(T)$ stand for the vector and tensor modes, respectively.
The function $b_{\rm K}(\eta,k)$ depends on the source of the vector and tensor modes (see Fig.~\ref{fig: bias K}).

For later purposes, we mention the non-magnetic scalar mode tidal effect. We adopt the linear alignment model, in which the galaxy ellipticity is linearly related to the real space density field:
\begin{align}
\gamma_{ij} &= b^{\rm scalar}_{\rm K}\left( \hat{k}_{i} \hat{k}_{j} - \frac{1}{3}\delta_{ij}\right) \delta_{\rm L}(\bm{k}) . \label{eq: gamma ij s}
\end{align}
Substituting Eq.~(\ref{eq: gamma ij s}) into Eq.~(\ref{eq: gamma pm 2}), we have
\begin{align}
{}_{\pm2}\gamma^{(S)} &= \frac{1}{2}b_{\rm K} \sin^{2}{\theta_{k}}\delta_{\rm L}(\bm{k}) e^{\pm 2i\phi_{k}} , \label{eq: gamma s}
\end{align}
for the scalar mode.

Using Eqs.~(\ref{eq: gamma s}), (\ref{eq: gamma vector}), and (\ref{eq: gamma tensor}), the expressions of the E- and B-modes are given by
\begin{align}
E^{(S)}(\bm{k}, \hat{\bm{n}}) &= \frac{1}{2}b^{\rm scalar}_{\rm K} \sin^{2}{\theta_{k}}\delta_{\rm L}(\bm{k}) ,  \label{eq: E scalar}\\
B^{(S)}(\bm{k}, \hat{\bm{n}}) &= 0 , \\
E^{(V)}(\bm{k}, \hat{\bm{n}}) &= \frac{1}{\sqrt{2}}b_{\rm K}(k)
\sin{\theta_{k}}\cos{\theta_{k}}
\sum_{\lambda=\pm1}\sigma^{(\lambda)}(\bm{k}) , \\
B^{(V)}(\bm{k}, \hat{\bm{n}}) &= -\frac{i}{\sqrt{2}}b_{\rm K}(k)
\sin\theta_{k} 
\sum_{\lambda=\pm1} \lambda \sigma^{(\lambda)}(\bm{k}) , \\
E^{(T)}(\bm{k}, \hat{\bm{n}}) &= \frac{1}{4}b_{\rm K}(k)
(1+\cos^{2}{\theta_{k}})
\sum_{\lambda=\pm2}h^{(\lambda)}(\bm{k}) , \\
B^{(T)}(\bm{k}, \hat{\bm{n}}) &= -\frac{i}{2}b_{\rm K}(k)
\cos\theta_{k} 
\sum_{\lambda=\pm2} \frac{\lambda}{2} h^{(\lambda)}(\bm{k}) , \label{eq: B tensor}
\end{align}
where the superscripts $(S)$, $(V)$, and $(T)$ stand for the scalar, vector, and tensor modes, respectively.

As the observable quantity, we focus on the three-dimensional power spectrum, which is defined as
\begin{align}
\Braket{X(\bm{k})Y^{*}(\bm{k}')} = (2\pi)^{3}\delta^{3}_{\rm D}(\bm{k}-\bm{k}') P_{XY}(\bm{k}) ,
\end{align}
where $X,\,Y = E$ or $B$ for the E- and B- mode power spectra. 
From Eqs.~(\ref{eq: E scalar})--(\ref{eq: B tensor}), we finally obtain
\begin{align}
P^{(S)}_{EE}(k, \mu) &= \frac{1}{4}\left(b^{\rm scalar}_{\rm K}\right)^{2} (1-\mu^{2})^{2}P_{\rm L}(k) , \label{eq: PEE scalar}\\
P^{(S)}_{BB}(k, \mu) &= 0 , \\
P^{(V)}_{EE}(k, \mu) & = \frac{1}{2} \left( b_{\rm K}(k) \right)^{2} (1-\mu^{2})\mu^{2} P_{\sigma}(k) ,  \label{eq: PEE vector}\\
P^{(V)}_{BB}(k, \mu) & = \frac{1}{2} \left( b_{\rm K}(k) \right)^{2} (1-\mu^{2}) P_{\sigma}(k) , \\
P^{(T)}_{EE}(k, \mu) & = \frac{1}{16} \left( b_{\rm K}(k) \right)^{2} (1+\mu^{2})^{2} P_{h}(k) , \\
P^{(T)}_{BB}(k, \mu) & = \frac{1}{4} \left( b_{\rm K}(k) \right)^{2} \mu^{2} P_{h}(k) , \label{eq: PBB tensor}
\end{align}
where we define $\mu = \cos{\theta}_{k}$. The linear matter power spectrum of the density field $\delta_{\rm L}$ is given by
\begin{align}
\Braket{\delta_{\rm L}(\bm{k})\delta^{*}_{\rm L}(\bm{k}')} = (2\pi)^{3}\delta^{3}_{\rm D}(\bm{k} - \bm{k}') P_{\rm L}(k) .
\end{align}
The non-vanishing $EB$ power spectrum appears in the chiral vector and tensor modes:
\begin{align}
P^{(V)}_{EB}(k, \mu) & = \frac{i}{2} \left( b_{\rm K}(k) \right)^{2} \mu \left( 1-\mu^{2}\right) \chi_{\sigma}(k) P_{\sigma}(k) , \\
P^{(T)}_{EB}(k, \mu) & = \frac{i}{8} \left( b_{\rm K}(k) \right)^{2} \mu \left( 1+\mu^{2}\right) \chi_{h}(k) P_{h}(k) , 
\end{align}
where we define a chiral parameter $\chi_{\sigma/h}(k)$ as
\begin{align}
\chi_{\sigma}(k) = \frac{P_{\sigma^{(+1)}}-P_{\sigma^{(-1)}}}{P_{\sigma}(k)} , \\
\chi_{h}(k) = \frac{P_{h^{(+2)}}-P_{h^{(-2)}}}{P_{h}(k)} .
\end{align}
The $EB$ spectrum is interesting probe for testing parity-violating theories.
Hereafter, we consider unpolarized cases $\chi_{\sigma/h}(k) = 0$.

\begin{figure}
\centering
\includegraphics[width=0.49\textwidth]{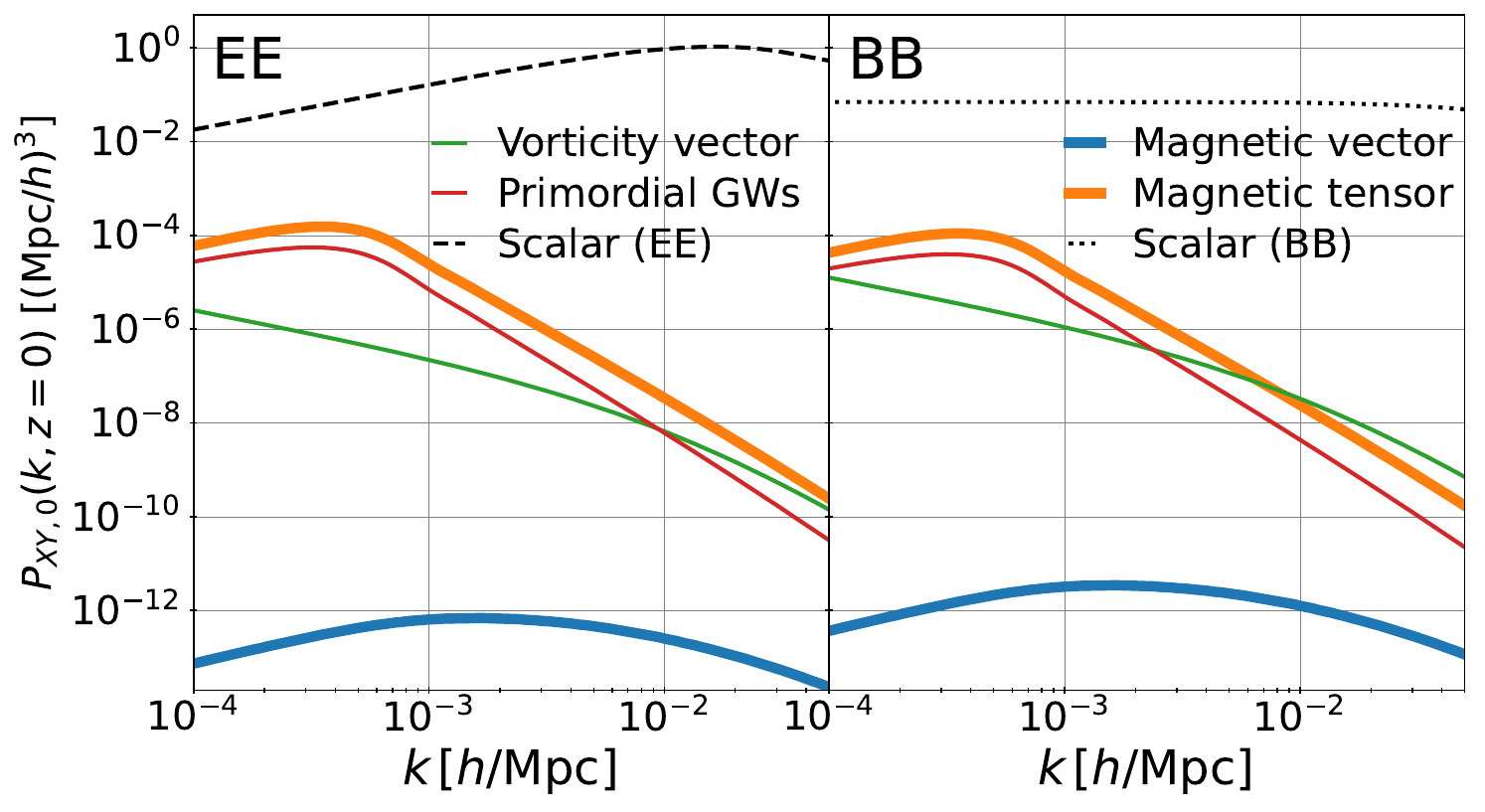}
\caption{E-mode (left) and B-mode (right) monopole power spectra at $z=0$ for various sources: the magnetic vector (blue), magnetic tensor (orange), vorticity vector (green) modes, and primordial GWs (red). To compute each case, we adopt $(B_{\lambda}, n_{B}) = (2\, {\rm nG}, -2.9)$, $(r_{\rm V}, n_{\rm V})= (0.01, 0)$ and $(r_{\rm T}, n_{\rm T}) = (0.03, -r_{\rm T}/8)$. Dashed and dotted lines represent, respectively, the E-mode power spectrum induced by the primary non-magnetic scalar mode given in Eq.~(\ref{eq: PEE scalar}) and the B-mode power spectrum induced by the one-loop contribution from the scalar mode, calculated based on Ref.~\cite{2023arXiv230916761C} (see Appendix~\ref{app: 1loop}).
}
\label{fig: ps all}
\end{figure}

We note that $P_{EE}^{(S)}$ is the non-magnetic scalar contribution, but affects the detectability of PMFs by forming the E-mode covariance of the Fisher matrix~(\ref{eq:Fish}). 
Moreover, at one-loop order, the density field can source the B mode, forming $P_{BB}^{(S)}$ (see Fig.~\ref{fig: ps all} and Appendix~\ref{app: 1loop}) and hence the B-mode covariance. We will also take them into account in the later Fisher matrix analysis.
For the vector and tensor power spectra $P_{\sigma/h}$, we adopt (see Sec.~\ref{sec: intro vector and tensor} in details)
\begin{align}
P_{\sigma}(k;B_{\lambda}, n_{B}) & = \left|\Pi^{(V)}\right|^{2} , \label{eq: Psigma mag vector} \\
P_{h}(k;B_{\lambda}, n_{B}) & = \left(6R_{\gamma}\ln\left( \frac{\eta_{\nu}}{\eta_{B}}\right) \right)^{2}
\left|\Pi^{(T)}\right|^{2} , \label{eq: Ph mag tensor} \\
P_{\sigma}(k;r_{\rm V}, n_{\rm V}) & =  \frac{2\pi^{2}}{k^{3}}r_{\rm V} \mathcal{A}_{\rm s}\left( \frac{k}{k_{\ast}} \right)^{n_{\rm V}} ,\\
P_{h}(k;r_{\rm T}, n_{\rm T}) & = \frac{2\pi^{2}}{k^{3}}r_{\rm T} \mathcal{A}_{\rm s} \left( \frac{k}{k_{\ast}} \right)^{n_{\rm T}},
\end{align}
for the magnetic vector, magnetic tensor, vorticity vector modes, and primordial GWs, respectively.

As a demonstration, we present the lowest order multipole, the monopole, of the various models in Fig.~\ref{fig: ps all}, although the non-vanishing quadrupole and hexadecapole are also observables. We define the monopole by
\begin{align}
P_{XX,0}(k) = \frac{1}{2}\int^{1}_{-1}{\rm d}\mu\, P_{XX}(k, \mu) .
\end{align}
In this plot, we set the model parameters as follows: $(B_{\lambda}, n_{B}) = (2\, {\rm nG}, -2.9)$ taken from the upper limit by Planck results~\cite{2016A&A...594A..19P}, $(r_{\rm V}, n_{\rm V})= (0.01, 0)$ roughly corresponding to the upper limit obtained by using the WMAP results~\cite{2012PhRvD..85d3009I}, and $(r_{\rm T}, n_{\rm T}) = (0.03, -r_{\rm T}/8)$ from the Planck results~\cite{2020A&A...641A...6P}.
Also, we calculate the scalar shape bias parameter $b^{\rm scalar}_{\rm K}$ by using the fitting formula~\cite{2021JCAP...04..041A}:
\begin{align}
b^{\rm scalar}_{\rm K} = \frac{0.09302 - 0.1289 b^{\rm E}_{1}}{1 + 0.3541 b^{\rm E}_{1}} , \label{eq: fit bK scalar}
\end{align}
where $b^{\rm E}_{1}$ is a linear density bias parameter. In this plot, we use the same linear density bias parameter as the HI galaxies in SKA2~\cite{2016ApJ...817...26B}:
\begin{align}
b^{\rm E}_{1} = c_{4}e^{c_{5}z} , \label{eq: bias SKA2}
\end{align}
with $c_{4} = 0.554$ and $c_{5} = 0.783$.
For comparison purposes, we show the contributions from the scalar mode to the E- and B-mode power spectra. While the leading-order effect of the scalar mode results only in the E mode, the one-loop order effect produces the non-vanishing B-mode contributions~(e.g., Refs.~\cite{2003PhRvD..68h3002H,2009ApJ...702..593S}). To compute the one-loop contribution to the B-mode spectrum, we exploit the effective-theory description of galaxy shape based on Ref.~\cite{2023arXiv230916761C}. See Appendix~\ref{app: 1loop} for details.

Recalling that all the primordial power spectrum are modeled by the power-law form, the characteristic feature observed in each power spectrum comes from the shape of each effective linear shape bias $b_{\rm K}$. The power spectrum of the magnetic tensor mode has a similar behavior to that of the primordial GWs as both $n_B = -2.9$ and $n_{\rm T} = -0.00375$ impose nearly scale invariance of their initial power spectra $P_h$ and also their $b_{\rm K}$ are exactly the same. Compared to the non-magnetic scalar power spectrum, the vector and tensor mode signals are suppressed at small scales because of the feature of the fossil effect, i.e., the absence of growth at late time.

We notice that the signal of the magnetic vector mode is 3--8 orders of magnitude smaller, depending on the scale, than those by other sources. We elaborate on the origin of this suppression as follows. 
First, to compute each spectrum in Fig.~\ref{fig: ps all}
we set the model parameters to the CMB limits; thus, the amplitude of each mode essentially reflects the amplitude of corresponding metric perturbation at around the recombination epoch. Indeed, there is a 2--4 orders of magnitude gap between the magnetic vector mode and the other threes in the metric perturbation, inducing a comparable gap in the IA.
Second, recalling the behavior of the effective linear shape bias in the magnetic vector mode (see the right panel in Fig.~\ref{fig: bias K}), the effective tidal bias asymptotically approaches zero more quickly than other modes at large scales. This asymptotic behavior leads to further suppression in the magnetic vector mode at large scales. The above two facts explain the behavior illustrated in Fig.~\ref{fig: ps all}.

\begin{figure*}
\centering
\includegraphics[width=\textwidth]{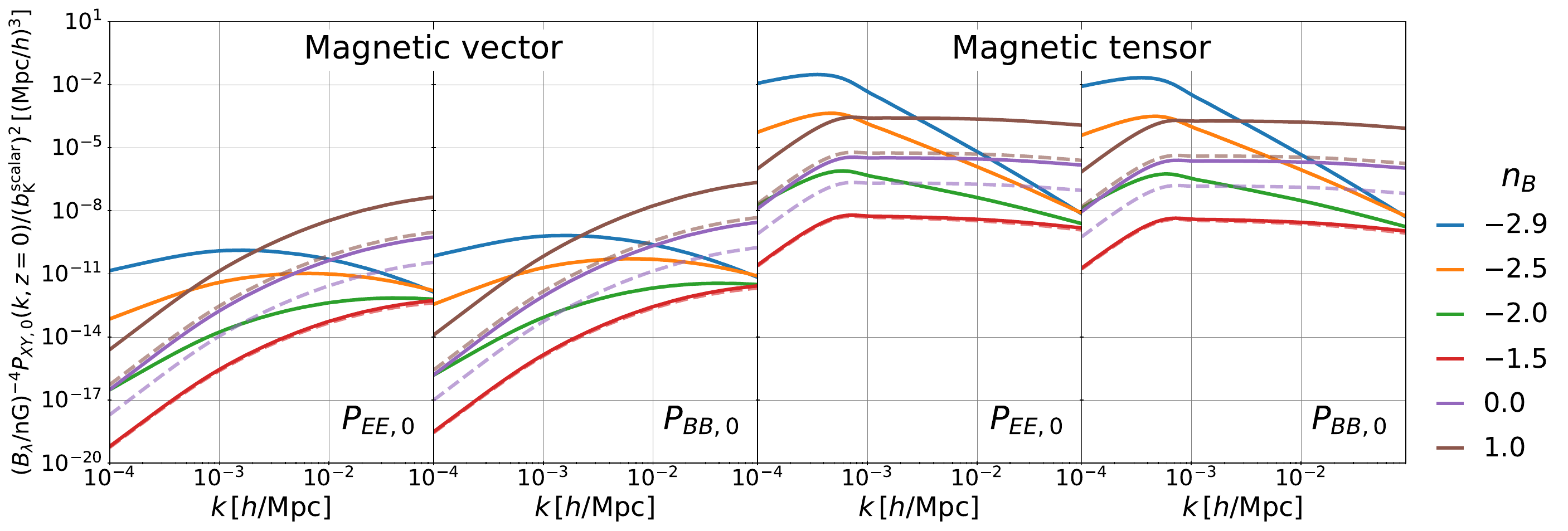}
\caption{E- and B-mode power spectra induced by PMFs normalized by $\left( B_{\lambda}/{\rm nG}\right)^{4}$ at $z=0$. Left two and right two panels, respectively, show the magnetic vector and magnetic tensor modes. Solid and dashed lines show the results in $B_{\lambda}=1.0$ and $10$ ${\rm nG}$, respectively. 
}
\label{fig: ps mag}
\end{figure*}

In Fig.~\ref{fig: ps mag}, we investigate the behavior of E- and B-mode power spectra of the magnetic modes by varying the model parameters $B_{\lambda}$ and $n_{B}$. Since we plot the signal normalized by $\left( B_{\lambda}/{\rm nG}\right)^{4}$, the solid and dashed lines overlap if the $P_{EE}$ and $P_{BB}$ scale as $\propto B^{4}_{\lambda}$. We indeed see this feature for $n_{B} \lesssim -1.5$, while their gap increases as $n_B$ gets larger than $-1.5$. This is because, for the blue tilted case, a convolution integral in the anisotropic stress~(\ref{eq: Pi ij}) becomes more sensitive to the ultra-violet magnetic cutoff $k_{\rm D}$ depending on $B_\lambda$ (see Eq.~(\ref{eq: kD})), and hence the power spectrum of the anisotropic stress no longer obeys a simple $B_\lambda^4$ scaling (see e.g. Refs.~\cite{2008PhRvD..78b3510F,2009MNRAS.396..523P,2011PhRvD..83l3533P}). Corresponding to the change of the impact of $k_{\rm D}$ at $n_B \sim -1.5$, the dependence of $P_{EE}$ and $P_{BB}$ on $n_B$ also changes, i.e., they decrease for small $n_B$ but start increasing as $n_B$ enlarges. As a consequence, they are minimized at $n_B \sim -1.5$. This unique feature straightforwardly determines the dependence of the detectability of $B_\lambda$ on $n_B$ as shown in Figs.~\ref{fig: Bl mag} and \ref{fig: Bl mag ideal}. The overall amplitude of the tensor mode is larger than that of the vector mode due to the prefactor $\left( 6R_{\gamma}\ln(\eta_{\nu}/\eta_{B})\right)^{2} \approx 2\times 10^{4}$. As we will demonstrate in the next section, the contribution from the tensor mode is an important source in constraining PMFs through the observation of the IA of galaxies.

\section{Fisher forecast\label{sec: Fisher}}

In this section, we discuss the constraining power of the E- and B-mode power spectra of the galaxy shape on the amplitude of PMFs based on the analytic model given in Eqs.~(\ref{eq: PEE vector})--(\ref{eq: PBB tensor}) with Eqs.~(\ref{eq: Psigma mag vector}) and (\ref{eq: Ph mag tensor}). To this end, we perform a Fisher matrix analysis. Following e.g. Ref.~\cite{2020ApJ...891L..42T}, we define the Fisher matrix for the parameter vector $\bm{\theta}$ as
\begin{align}
F_{ij} &= \frac{V}{(2\pi)^{2}}
\int^{k_{\rm max}}_{k_{\rm min}}k^{2}{\rm d}k \,
\int^{1}_{-1}{{\rm d}\mu} \,
\notag \\
& \times
\sum_{a,b=EE, BB} 
\left( \frac{\partial P_{a}}{\partial \theta_{i}} \right)
\left[ {\rm cov}^{-1} \right]_{ab}
\left( \frac{\partial P_{b}}{\partial \theta_{j}} \right) , \label{eq:Fish}
\end{align}
where the quantity $V$ represents the survey volume.
The covariance matrix ${\rm cov}_{ab}$ is given by
\begin{align}
{\rm cov}_{ab} = 2\left( P_{a} + \frac{\sigma^{2}_{\gamma}}{n_{\rm gal}}\right)^{2} \delta_{a,b} .
\end{align}
with $\sigma_{\gamma}$ and $n_{\rm gal}$ being the root-mean-square of the galaxy's ellipticity and the galaxy number density, respectively.

Our analysis examines the constraints on the amplitude of PMFs, $B_{\lambda}$ with a fixed spectral index $n_{B}$. In this case, the expression of the Fisher matrix is reduced to
\begin{align}
F(B_{\lambda,{\rm fid}}) & = \frac{V}{2(2\pi)^{2}}
\int^{k_{\rm max}}_{k_{\rm min}}k^{2}{\rm d}k \,
\int^{1}_{-1}{{\rm d}\mu} \,
\notag \\
& \times 
\sum_{a=EE, BB}
\left( \left.
\frac{\partial P_{a}}{\partial B_{\lambda}} 
\frac{1}{P_{a} + \sigma^{2}_{\gamma}/n_{\rm gal}}
\right|_{B_{\lambda} = B_{\lambda,{\rm fid}}}
\right)^{2}
,
\end{align}
with $B_{\lambda,{\rm fid}}$ being the fiducial value the parameter. The size of the expected error on the PMF strength is given by $\sigma(B_{\lambda, {\rm fid}}) = \sqrt{F^{-1}(B_{\lambda,{\rm fid}})}$. PMFs whose strength exceeds the size of error are detectable at the 1$\sigma$ level; thus, a minimum detectable value of the PMF strength $B_{\lambda,{\rm min}}$ is given by a solution of the equation
\begin{align}
\sigma(B_{\lambda, {\rm min}}) = B_{\lambda, {\rm min}} . \label{eq: min Blambda}
\end{align}

Throughout the analysis, we incorporate the E-mode power spectrum induced by the scalar mode and the B-mode power spectrum induced by the vector and tensor modes and the one-loop contribution from the scalar mode into the Fisher analysis. In computing the one-loop B-mode spectrum, we follow Ref.~\cite{2023arXiv230916761C} and use the effective-theory description of galaxy shape (see Appendix~\ref{app: 1loop} for details). We set $k_{\rm min} = 2\pi V^{-1/3}$, $k_{\rm max} = 0.1 \, {\rm Mpc}/h$, and $\sigma_{\gamma} = 0.3$.
The scalar shape bias parameter $b^{\rm scalar}_{\rm K}$ is calculated by using the fitting formula~\cite{2021JCAP...04..041A} given in Eq.~(\ref{eq: fit bK scalar}).

\begin{figure}
\centering
\includegraphics[width=0.49\textwidth]{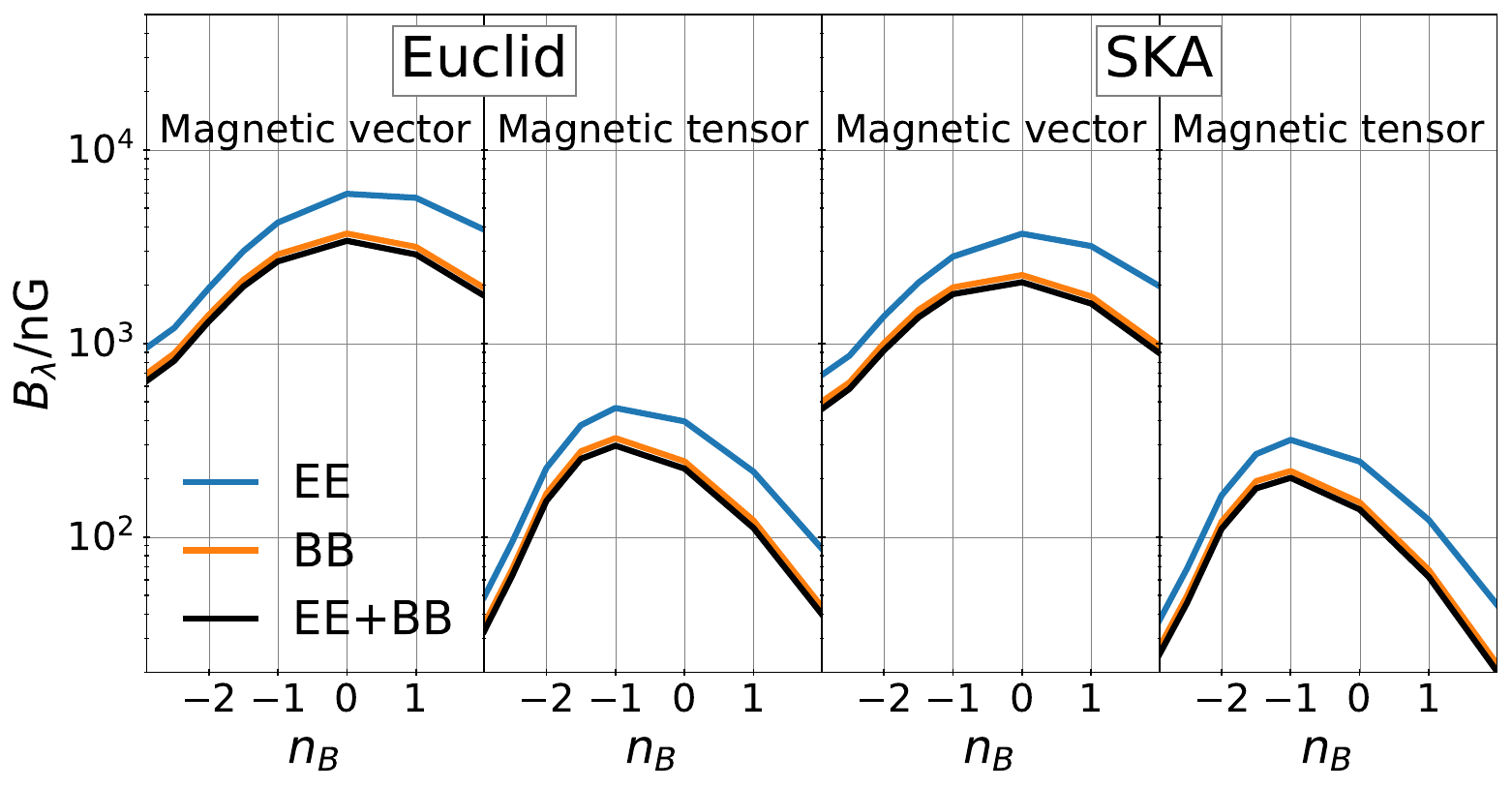}
\caption{Minimum detectable value for the Euclid spectroscopic survey (left two panels) and SKA HI galaxy surveys (right two panels) obtained through the magnetic vector mode and through the magnetic tensor mode as indicated. Blue, orange, and black lines, respectively, show the results obtained by E-mode power spectrum alone, by B-mode power spectrum alone, and by combining E- and B-mode power spectra, respectively. We note that the orange and black lines, corresponding to the results from B mode alone and from both E and B modes, respectively, almost overlap, showing that the B mode gives a much larger gain than the E mode. For 
visualization purposes, we multiply the B-mode results by a factor of 1.12.}
\label{fig: Bl mag}
\end{figure}

\begin{figure*}
\centering
\includegraphics[width=\textwidth]{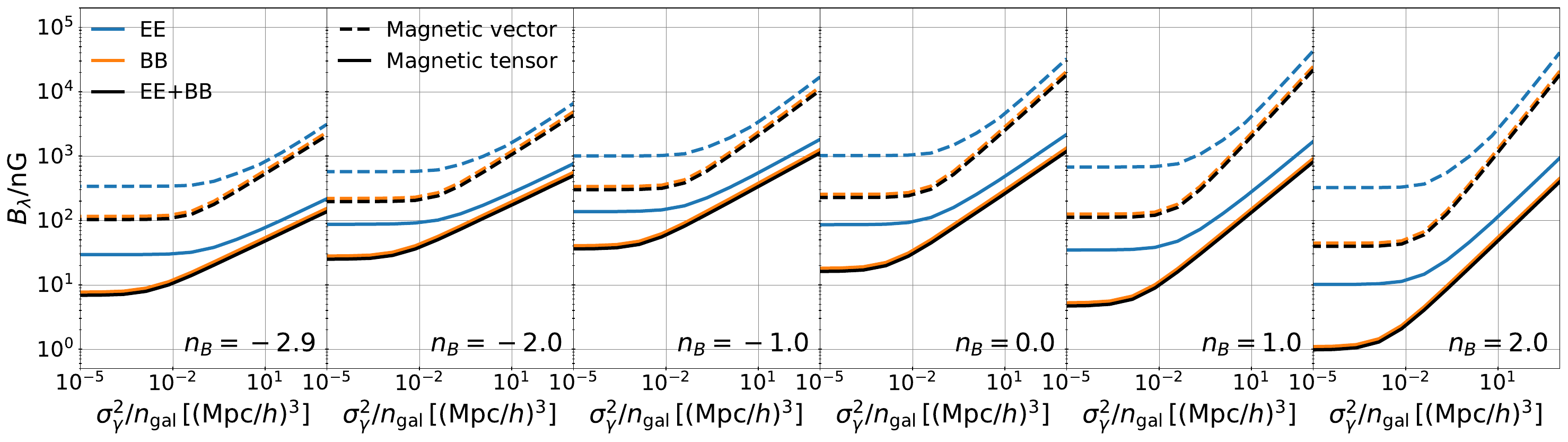}
\caption{Minimum detectable value for the idealistic surveys by varying $\sigma^{2}_{\gamma}/n_{\rm gal}$, obtained through the magnetic vector mode (dashed lines) and through the magnetic tensor mode (solid lines). We present the results obtained by E-mode power spectrum alone (blue), by B-mode power spectrum alone (orange), and by combining E- and B-mode power spectra (black). The orange and black lines, corresponding to the results from B mode only and from both E and B modes, respectively, almost overlap, showing that the B mode gives a much bigger gain than the E mode. For 
visualization purposes, we multiply the B-mode results by a factor of 1.12.}
\label{fig: Bl mag ideal}
\end{figure*}

We first demonstrate the expected minimum detectable value for Euclid spectroscopic survey~\cite{2011arXiv1110.3193L} and Square Kilometre Array (SKA)~\cite{2020PASA...37....7S}. The galaxy redshift surveys by Euclid and SKA will, respectively, observe the H$\alpha$ emitter over redshifts 0.9 to 1.8 and HI galaxies over redshifts 0.23 to 1.81. 
Although IA has not yet been detected for emission line galaxies (ELGs)~\cite{2006MNRAS.367..611M,2011MNRAS.410..844M,2018PASJ...70...41T,2022ApJ...924L...3T}, Ref.~\cite{2021ApJ...917..109S} recently proposed an optimal estimator to determine the IA of halos using ELGs. We suppose that the power spectra related to the IA can be measured with the optimal estimator, and that all observed ELGs are, therefore, an ideal tracer of the halo shape. 
We use the survey specifications in Table~3 of Ref.~\cite{2020A&A...642A.191E} for the Euclid and Table~1 of Ref.~\cite{2015aska.confE..24B} for SKA. We combine all the redshift bins by $\sigma(B_{\lambda}) = 1/\sqrt{\sum_{i}\left.F(B_{\lambda})\right|_{z=z_{i}}}$ with the subscript $i$ being the label of the redshift bins, and then numerically solve Eq.~(\ref{eq: min Blambda}). In Fig.~\ref{fig: Bl mag}, we present the minimum detectable value for Euclid and SKA (see Appendix~\ref{app: others} for the same analysis but for the vorticity vector mode and primordial GWs). As expected from Fig.~\ref{fig: ps mag}, the magnetic tensor mode provides stronger constraints on the PMFs strength than the magnetic vector mode. It is also apparent from Fig.~\ref{fig: Bl mag} that the B-mode power spectrum places stronger constraints than the E-mode power spectrum due to the absence of the sizable non-magnetic scalar contribution in the covariance matrix. We note that, since the lower wavenumber limit of the integration range in the Fisher matrix is $k_{\rm min} = 2\pi V^{-1/3} \approx 0.001 h/{\rm Mpc}$ under the survey specification of Euclid/SKA, a unique $n_{B}$ dependence of $B_{\lambda, {\rm min}}$ seen in Fig.~\ref{fig: Bl mag}; namely, maximized at $n_B \sim -1$, is due to a unique $n_{B}$ dependence of $P_{EE}$ and $P_{BB}$ above this scale; namely, minimized at $n_B \sim -1$, (see Fig.~\ref{fig: ps mag}). One can find from Fig.~\ref{fig: Bl mag} that PMFs with $B_\lambda \sim 30$--$300 \, {\rm nG}$ would be measurable by a Euclid- or SKA-level B-mode survey.

To capture PMFs with $B_\lambda = {\cal O}(1 \, {\rm nG})$, what specific level of survey should be aimed at? To figure out this, we compute the minimum detectable value $B_{\lambda, \rm min}$ by varying the shot noise contribution to the covariance matrix $\sigma^{2}_{\gamma}/n_{\rm gal}$ (see Appendix~\ref{app: others} for the same analysis but for the vorticity vector mode and primordial GWs). We then set $V = 1\, ({\rm Gpc}/h)^{3}$ and $z=1.0$. We also use the same bias parameter as the HI galaxies in SKA2 given in Eq.~(\ref{eq: bias SKA2}). The choice of these parameters does not qualitatively change the results. Fig.~\ref{fig: Bl mag ideal} describes our results, again showing that the B-mode information induced by the tensor mode is most powerful to measure $B_\lambda$.
Both for the E-mode and B-mode cases, decreasing the shot noise level causes the dominance of the non-magnetic scalar signal in the covariance and hence the saturation of $B_{\lambda, \, \rm min}$ around $\sigma^{2}_{\gamma}/n_{\rm gal} \sim 10^{-2}$. We finally find that the saturated value of $B_{\lambda, \, \rm min}$ reaches $\mathcal{O}(1\, {\rm nG})$--$\mathcal{O}(10\, {\rm nG})$, depending on the spectral index. To archive this minimum detectable value, the B-mode power spectrum plays a major role and still an interesting probe even in the presence of the one-loop non-magnetic scalar contribution to the B mode.

\section{Summary\label{sec: summary}}

In recent years, there has been growing attention on primordial magnetic fields (PMFs) as a strong candidate for explaining the origin of observed large-scale magnetic fields, including void regions. While the statistical properties of PMFs have been constrained by the recent cosmological observations such as the cosmic microwave background anisotropies, this paper has focused on the intrinsic alignments (IAs) of galaxies as a complementary new observational probe, aiming to delve into the nature of PMFs. The metric perturbations of the long-wavelength vector and tensor modes are known to induce the local tidal gravitational fields~\cite{2014PhRvD..89h3507S}. 
Through the observations of the intrinsic galaxy shapes, we have paved the way to detect the vector and tensor modes sourced by the anisotropic stress of PMFs in the early Universe.

We have shown the relation between the anisotropic stress of PMFs and the IA of the galaxies.
Considering up to the leading-order contributions, while the scalar mode only produces the cosmic shear E-mode, the vector and tensor modes produce both E- and B-modes.
Hence, the B-mode signal can be a good probe to search for magnetically induced vector and tensor modes once we properly take into account the one-loop scalar contribution to the B mode. Assuming that the statistical properties of PMFs are given by a power-low type power spectrum, which includes two parameters: the amplitude of PMFs $B_{\lambda}$ and the spectral index $n_{B}$, we demonstrated the E- and B-mode power spectra of the galaxy shape induced by PMFs. Due to the convolution and small-scale cut-off inherent in the anisotropic stress of PMFs, we found that the slopes of the E- and B-mode spectra do not change for $n_{B}>-1.5$ as $n_{B}$ is increased, but only their amplitudes vary with $n_{B}$.

Based on our analytical model of the E- and B-mode spectra, we have performed the Fisher analysis to estimate the minimum detectable value of the PMF strength, defined in Eq.~(\ref{eq: min Blambda}), for a fixed spectral index. We first examined the minimum detectable value assuming the galaxy redshift survey by Euclid and SKA. In this case, we found that the minimum detectable value of $B_{\lambda}$ reaches about $30$--$300\, {\rm nG}$, depending on $n_{B}$, which is weaker than the upper limit obtained by the recent CMB observations. To investigate the detecting power of the IA observations in spectroscopic surveys, we further performed the Fisher matrix analysis by varying the shot noise term as a free parameter.
We found that a minimum detectable value can reach $\mathcal{O}(1\,{\rm nG})$--$\mathcal{O}(10\,{\rm nG})$, depending on $n_{B}$, which is almost comparable to the current CMB limits, and that the B-mode spectrum still plays a crucial role in achieving this even in the presence of the non-magnetic scalar contribution to the B mode spectrum.
The currently planned galaxy redshift surveys would provide weaker constraints on PMFs than the CMB observations. However, the observations of the galaxy IAs would become increasingly important as a complementary probe to understand the nature of PMFs.

This paper has focused on the auto power spectra of the cosmic shear E- and B-modes induced by PMFs. However, as the anisotropic stress of PMFs also induces the density fluctuations~\cite{2010PhRvD..81d3517S}, we would observe a non-vanishing signal in the cross-correlation between density fields and galaxy shapes. Adding this information to the present analysis would improve the constraint on PMFs. An interesting future challenge is to probe PMFs, making comprehensive use of the available information on the galaxy density field and its shape.

We have carried out our analysis with a spectroscopic survey in mind. When considering an analysis based on the two-dimensional angular power spectrum for a photometric survey (e.g. Ref.~\cite{2023arXiv230908653P}), we expect that the impact of the shot noise on the covariance is reduced due to the larger number density of galaxies than spectroscopic surveys. We leave detailed comparisons of the detecting power on PMFs between two-dimensional angular power spectrum with three-dimensional power spectrum for an intriguing future work.

\begin{acknowledgments}
SS and KA are supported by JSPS Overseas Research Fellowships.
This work is supported by the Japan Society for the Promotion of Science (JSPS) KAKENHI Grant Nos. JP23K19050 (SS), JP20H05859 (MS) and JP23K03390 (MS).
KA also acknowledges support from Fostering Joint International Research (B) under Contract No.~21KK0050.
TO acknowledges support from the Ministry of Science and Technology of Taiwan under Grants No. MOST 111-2112-M-001-061- and NSTC 112-2112-M-001-034- and the Career Development Award, Academia Sinica (AS-CDA-108-M02) for the period of 2019-2023.
\end{acknowledgments}
\appendix

\section{Intrinsic alignments from vector and tensor modes\label{app: fossil}}

In this Appendix, based on Ref.~\cite{2014PhRvD..89h3507S}, we solve the equation of motion of a matter particle in a local frame in the presence of the local tidal effect, and then derive the density fields induced by the cross-talk between the long- and short-wavelength modes. From the expression for the second-order density fields, we show the linear shape bias of the IA of galaxies from the vector and tensor modes.

Our starting point is the expression of the tidal field induced by the long-wavelength vector and tensor modes using conformal Fermi normal coordinate~\cite{2014PhRvD..89h3507S}:
\begin{align}
\tau_{ij}(\eta,k_{\rm L}) & = - \frac{1}{2a}\left( a h'_{ij}(\eta,k_{\rm L}) \right)' ,
\end{align}
where we work in the synchronous gauge. It is useful to decompose the tidal tensor into the time-dependent part and initial perturbation part:
\begin{align}
\tau_{ij}(\eta,k_{\rm L}) = \mathcal{T}_{\tau}(\eta,k_{\rm L}) \xi_{{\rm ini}\, ij}(k_{\rm L}) .
\end{align}
where $\xi = \sigma$ and $h$ for the vector and tensor modes, respectively.
The function $\mathcal{T}_{\tau}(\eta,k_{\rm L})$ is given by
\begin{align}
\mathcal{T}_{\tau}(\eta,k_{\rm L}) & = - \frac{k_{\rm L}}{2a}\left( a\mathcal{T}^{(V)}(\eta,k_{\rm L})\right)' ,
\end{align}
for the vector mode, and
\begin{align}
\mathcal{T}_{\tau}(\eta,k_{\rm L}) & = - \frac{1}{2a}\left( a\mathcal{T}^{(T)\prime}(\eta,k_{\rm L})\right)' ,
\end{align}
for the tensor mode.

In the presence of the long-wavelength tidal tensor $\tau_{ij}$, the equation of motion of a matter particle in the local frame becomes
\begin{align}
\bm{x}'' + \mathcal{H}\bm{x}' &= -\bm{\nabla}_{x}\left( \phi_{\rm s} + \frac{1}{2}\tau_{ij}x^{i}x^{j}\right) , \label{eq: app EoM} \\
\nabla^{2}_{x}\phi_{\rm s} &= 4\pi G a^{2} \bar{\rho}_{\rm m} \delta , \label{eq: app Poisson}
\end{align}
where a prime denotes a derivative with respect to the conformal time $\eta$.
To solve the equation of motion, we employ the Lagrangian perturbation formalism. 
The Lagrangian description relates the initial Lagrangian position for the fluid element $\bm{q}$ to the Eulerian position at conformal time $\eta$ through the displacement field $\bm{\Psi}(\eta,\bm{q})$:
\begin{align}
\bm{x}(\eta,\bm{q}) = \bm{q} + \bm{\Psi}(\eta,\bm{q}) . \label{eq: x to q}
\end{align}

Substituting Eq.~(\ref{eq: x to q}) into Eq.~(\ref{eq: app EoM}), the equation for $\bm{\Psi}$ at the first order is given by
\begin{align}
\bm{\Psi}'' + \mathcal{H}\bm{\Psi}' = -\bm{\nabla}_{q}\left( 
\phi_{\rm s}(\bm{q}) + \frac{1}{2}\tau_{ij}q^{i}q^{j}\right) . \label{eq: EoM 1st}
\end{align}
We split the displacement field into the long- and short-wavelength mode contributions:
\begin{align}
\bm{\Psi} = \bm{\Psi}^{(s)} + \bm{\Psi}^{(l)} .
\end{align}
The evolution equation of each displacement field is given by
\begin{align}
\Psi^{(s)\prime\prime}_{i} + \mathcal{H} \Psi^{(s)\prime}_{i} &= -\frac{\partial \phi_{\rm s}(\bm{q})}{\partial q_{i}} , \label{eq: 1st short}\\
\Psi^{(l)\prime}_{i} + \mathcal{H} \Psi^{(l)\prime}_{i} &
= - \mathcal{T}_{\tau}(\eta,k_{\rm L})\xi_{{\rm ini}\, ia}q_{a}
, \label{eq: 1st long}
\end{align}

Using Poisson equation~(\ref{eq: app Poisson}), the solution of Eq.~(\ref{eq: 1st short}) is given by
\begin{align}
\Psi^{(s)}_{i}(\eta,\bm{q})
= - D(\eta) \partial^{-2}_{q}\frac{\partial}{\partial q_{i}}\delta^{(1)}(\eta_{0},\bm{q}) , \label{eq: app sol s 1}
\end{align}
where the quantity $\eta_{0}$ is the conformal time at the present time.
The linear growth factor $D(\eta)$ satisfies
\begin{align}
D''(\eta) + \mathcal{H}D'(\eta) - 4\pi G a^{2} \bar{\rho}_{\rm m} D(\eta) = 0 .
\end{align}
We adopt the normalization conditions $D(\eta_{0}) = 1$.
The solution of Eq.~(\ref{eq: 1st long}) is given by
\begin{align}
\Psi^{(l)}_{i} &= -\beta(\eta,k_{\rm L}) \xi_{{\rm ini}\, ia}q_{a} , \label{eq: app sol l 1}
\end{align}
where we define 
\begin{align}
\beta(\eta,k_{\rm L})
&= 
- \int^{\eta}_{0}\frac{{\rm d}\eta'}{a(\eta')}
\int^{\eta'}_{0}\,  a(\eta'') \mathcal{T}_{\tau}(\eta'',k_{\rm L})  {\rm d}\eta'' .
\end{align}

Next, we solve the equation for $\bm{\Psi}$ by considering only the coupling between long- and short-wavelength modes. We start from taking the divergence of Eq.~(\ref{eq: app EoM}) with respect to $\bm{q}$:
\begin{align}
\Psi^{(sl)\prime\prime}_{a,a} + \mathcal{H} \Psi^{(sl)\prime}_{a,a}
= &
- \nabla^{2}_{x}\phi_{\rm s}
- \Psi^{(s)}_{b,a} \tau_{ab}
- \Psi^{(l)}_{b,a} \partial_{x_{b}} \partial_{x_{a}}\phi_{\rm s} ,
\end{align}
where a comma stands for the derivative with respect to the Lagrangian coordinate. 
We note that the first term on the right-hand side involves the coupling of short- and long-wavelength modes through the chain rule of spatial difference: $\frac{\partial}{\partial x_{i}} = \left( J^{-1}\right)_{ji}\frac{\partial}{\partial q_{j}}$ with the Jacobian matrix $J_{ij} = \frac{\partial x_{i}}{\partial q_{j}} = \delta_{ij} + \Psi_{i,j}$. Then, the second order equation for $\bm{\Psi}^{(sl)}$ is given by
\begin{align}
& \Psi^{(sl)\prime\prime}_{a,a} + \mathcal{H} \Psi^{(sl)\prime}_{a,a} - 4\pi G a^{2} \bar{\rho}_{\rm m} \Psi^{(sl)}_{a,a}
\notag \\
& \qquad 
= \mathcal{T}_{\tau}(\eta,k_{\rm L}) D(\eta)
\partial^{-2}_{q}\frac{\partial^{2}\delta^{(1)}(\eta_{0},\bm{q})}{\partial q_{a}\partial q_{b}}
\xi_{{\rm ini}\, ab} \label{eq: app sol ls 2}
\end{align}
The solution of this equation is given by
\begin{align}
\Psi^{(sl)}_{a,a} = 
D^{(sl)}(\eta,k_{\rm L})\partial^{-2}_{q}\frac{\partial^{2}\delta^{(1)}(\eta_{0})}{\partial q_{a}\partial q_{b}}
\xi_{{\rm ini}\, ab}(k_{\rm L})
\end{align}
where the function $D^{sl}(\eta,k_{\rm L})$ satisfies 
\begin{align}
& D^{(sl)\prime\prime} + \mathcal{H} D^{(sl)\prime} - 4\pi G a^{2} \bar{\rho}_{\rm m} D^{(sl)}
= D(\eta) \mathcal{T}_{\tau}(\eta, k_{\rm L})
\end{align}

From Eqs.~(\ref{eq: app sol s 1}), (\ref{eq: app sol l 1}), and (\ref{eq: app sol ls 2}), we obtain the Eulerian density field up to the second order:
\begin{align}
\delta(\bm{x}) & = \delta^{(s)}(\bm{x}) + \delta^{(sl)}(\bm{x}),
\end{align}
where we define
\begin{align}
\delta^{(s)}(\bm{x}) &= -\Psi^{(s)}_{a,a}(\bm{x}) , \notag  \\
& = D(\eta) \delta^{(1)}(\eta_{0}, \bm{x}) , \\
\delta^{(sl)}(\bm{x})
&= 
\Psi^{(s)}_{a,ab}(\bm{x})\Psi^{(l)}_{b}(\bm{x}) - \Psi^{(sl)}_{a,a}(\bm{x})
\notag \\
& \quad 
+ \Psi^{(s)}_{a,a}(\bm{x})\Psi^{(l)}_{b,b}(\bm{x}) + \Psi^{(s)}_{a,b}(\bm{x})\Psi^{(l)}_{b,a}(\bm{x}) , \\
&=
\xi_{{\rm ini}\, ab}(k_{\rm L})
\Biggl[
\beta(\eta,k_{\rm L}) x_{a} \partial_{b}
\notag \\
&
+ \left( - \frac{D^{(sl)}(\eta,k_{\rm L})}{D(\eta)} 
+ \beta(\eta,k_{\rm L}) \right) \partial^{-2}\partial_{a}\partial_{b}
\Biggr]
\delta^{(1)}(\bm{x}) . \label{eq: app 2nd order delta}
\end{align}

We use the same ansatz in Ref.~\cite{2014PhRvD..89h3507S} which assumes that the second term in the square brackets in Eq.~(\ref{eq: app 2nd order delta}) induces the IAs, since this term represents the growth of the density perturbation in a local region by the coupling between the long- and short-wavelength tidal fields while the first term corresponds to the displacement induced by the long-wavelength tidal field, which should have no effect on local physics.
We also assume that the alignment from the vector/tensor tidal fields has the same scaling as the second order density induced by the scalar tidal fields. 
According to this ansatz, the expression of the intrinsic galaxy shape is given by
\begin{align}
\gamma_{ij} &= b_{\rm K}(\eta,k_{\rm L})\xi_{{\rm ini}\, ij}, \\
b_{\rm K}(\eta,k_{\rm L}) & \equiv \frac{7}{4} \left( - \frac{D^{(sl)}(\eta,k_{\rm L})}{D(\eta)}   + \beta(\eta,k_{\rm L}) \right) b^{\rm scalar}_{\rm K} . \label{eq: bK}
\end{align}
where $b_{\rm K}^{\rm scalar}$ is the scalar linear shape bias. 
The factor $7/4$ comes from the conversion from the second-order density field to the galaxy intrinsic shape in the scalar mode case~\cite{2014PhRvD..89h3507S,2023PhRvD.107f3531A}.

\section{One loop contributions to B mode power spectrum\label{app: 1loop}}
Here we give a brief explanation on what is assumed to compute the one-loop correction to the $B$-mode auto power spectrum.
Recently the perturbation theory of the IAs with the effective theory considerations has been formulated in both Eulerian and Lagrangian ways~\cite{2020JCAP...01..025V,2023arXiv230916761C}.
In this paper we employ the LPT-based calculation of the one-loop power spectrum developed in Ref.~\cite{2023arXiv230916761C}.

As the one-loop power spectrum involves the linear, quadratic, and cubic fields, we need to introduce up to the cubic shape bias parameters, which in general consists of one linear, three quadratic, two cubic free parameters to compute the one-loop correction to the shape power spectrum.
However, with the comparison of the $N$-body halo shapes, Ref.~\cite{2023JCAP...08..068A} showed that the values of the higher order (Eulerian) shape bias parameters are well approximated by the coevolution prediction~\cite{2018JCAP...07..030S}.
In other words, the halo shape field is well described by the Lagrangian tracer of the initial tidal field advected to its final position by the large-scale bulk flow.
Hence the following model can be used in lieu of the full model for the one-loop power spectrum:
\begin{align}
    \gamma_{ij}({\bf k}) = \int {\rm d}^3 {\bf q}~
    \gamma^{\rm L}_{ij}({\bf q})
    e^{i{\bf k}\cdot({\bf q}+{\boldsymbol \Psi}({\bf q}))},
\end{align}
with
\begin{align}
    \gamma^{\rm L}_{ij}({\bf q})
    =
    b^{\rm L}_{\rm K} K_{ij}({\bf q}) 
    [1 + \delta_g^{\rm L}({\bf q})],
    \label{eq: Lagrangian shape}
\end{align}
where $b_{\rm K}^{\rm L}$ is the Lagrangain linear shape bias and $\delta_g^{\rm L}({\bf q})$ is the Lagrangian galaxy density field.
Note that since the galaxy shapes are always observed with galaxies they are naturally density-weighted quantities.

In order to compute the one-loop correction, we can also assume the linear bias description for the galaxy density field since the quadratic bias fields in the density in Eq.~\eqref{eq: Lagrangian shape} give rise to a reparametrization of the linear shape bias parameter. After all, with these assumptions, the free bias parameters we have to include to the one-loop calculation are the linear shape and density bias parameters: $b_{\rm K}^{\rm L}$ and $b_1^{\rm L}$.
The Lagrangian linear shape bias is the same as the Eulerian one, $b_{\rm K}^{\rm scalar} = b_{\rm K}^{\rm L}$, since the tidal field does not induce the volume distortion at first order, while the Lagrangian linear density bias is related to the Eulerian one as $b_1^{\rm E} = b_1^{\rm L} + 1$.
Using these relations we can compute the one-loop correction to the shape power spectrum given the values of $b_{\rm K}^{\rm scalar}$ and $b_1^{\rm E}$.

\section{Fisher forecast for the vorticity vector mode and primordial GWs\label{app: others}}

\begin{figure}
\centering
\includegraphics[width=0.49\textwidth]{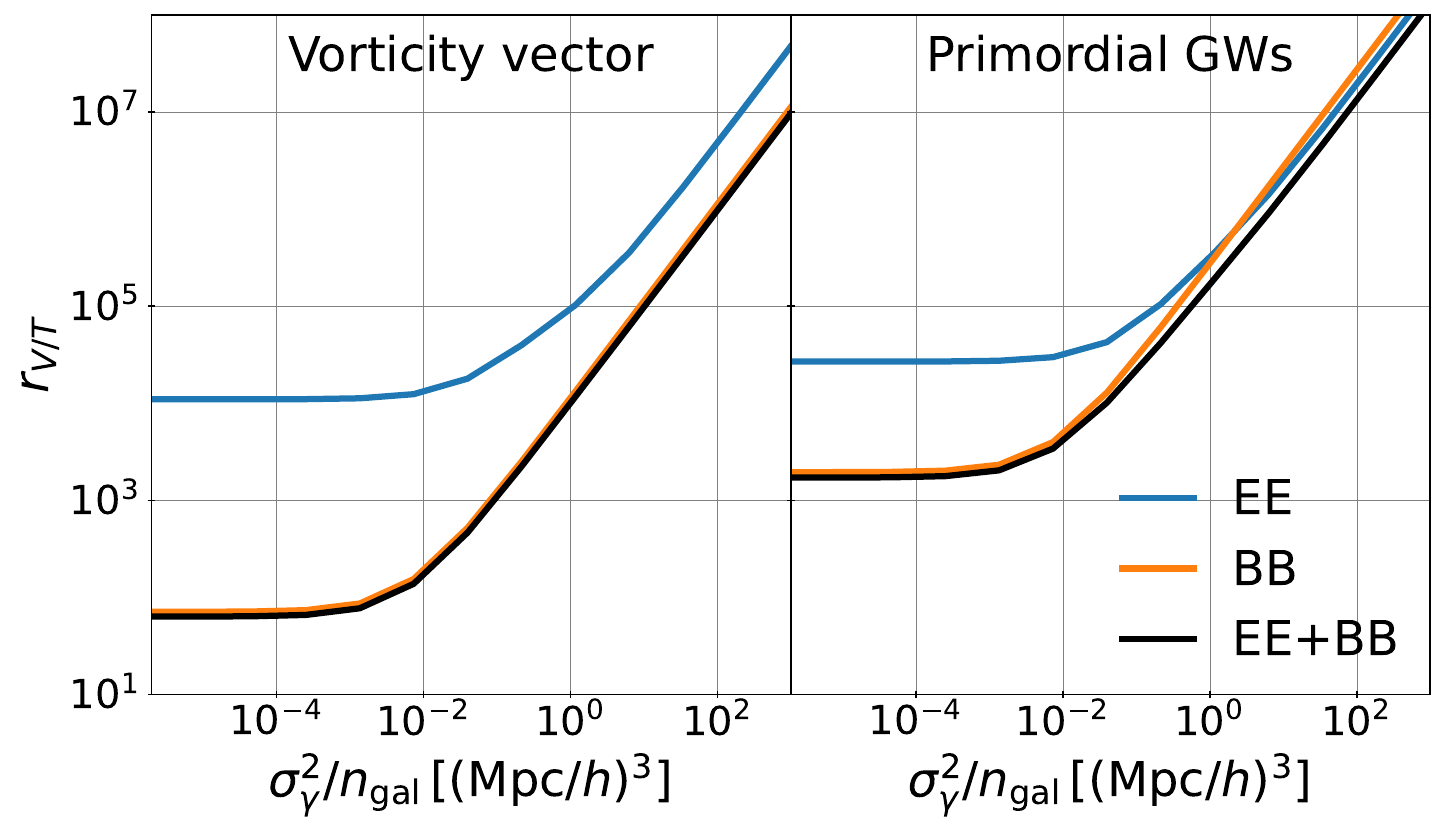}
\caption{Same as Fig.~\ref{fig: Bl mag ideal} but for the vorticity vector mode and primordial GWs.}
\label{fig: primordial V and T}
\end{figure}

As we are interested in the detectability of the PMFs through observations of the galaxy shape, we have focused on the vector and tensor modes induced by PMFs in the main text. For reference purposes, this appendix provides the Fisher forecast based on the same analysis as done in Sec.~\ref{sec: Fisher} but we consider other vector and tensor sources: the vorticity vector mode and the primordial GWs.

We define the Fisher matrix for a fixed spectral index $n_{V} = n_{T} = 0$ by
\begin{align}
F(r_{X, {\rm fid}}) & = \frac{V}{2(2\pi)^{2}}
\int^{k_{\rm max}}_{k_{\rm min}}k^{2}{\rm d}k \,
\int^{1}_{-1}{{\rm d}\mu} \,
\notag \\
& \times 
\sum_{a=EE, BB}
\left( \left.
\frac{\partial P_{a}}{\partial r_{X}} 
\frac{1}{P_{a} + \sigma^{2}_{\gamma}/n_{\rm gal}}
\right|_{r_{X} = r_{X, {\rm fid}}}
\right)^{2}
,
\end{align}
where $X=V$ and $T$ for the vorticity vector mode and the primordial GWs, respectively. The size of the expected error on the amplitude of the vector/tensor modes is given by $\sigma(r_{X, {\rm fid}}) = \sqrt{F^{-1}(r_{X, {\rm fid}})}$. We here evaluate the minimum detectable value $r_{X,{\rm min}}$ by solving
\begin{align}
\sigma(r_{X,{\rm min}}) = r_{X,{\rm min}} .
\end{align}

\begin{table}
\begin{tabular}{c|c|c}
Euclid / SKA & Vorticity vector $10^{-5}r_{\rm V}$
& Primordial GWs $10^{-5}r_{\rm T}$ 
\\
 \hline
EE
& 2.95 / 0.84
& 4.45 / 1.45
\\
BB
& 0.631 / 0.171
& 5.35 / 1.66
\\
EE$+$BB
& 0.615 / 0.167
& 3.19 / 1.04
\end{tabular}
\caption{Minimum detectable value for the spectroscopic survey in Euclid (left values) and SKA (right values). For the spectral index of the vorticity vector mode and primordial GWs, we set $n_{V} = n_{T} = 0$.}
\label{tab: Euclid / SKA vector and tensor}
\end{table}

Table~\ref{tab: Euclid / SKA vector and tensor} shows the results for the Euclid spectroscopic survey and SKA HI galaxy survey. We see from this that the E-mode (B-mode) power spectrum can capture smaller $r_T$ ($r_V$) than the B-mode (E-mode) one. This can happen in surveys where the covariance is dominated by the shot noise as the Euclid and SKA. For the primordial GW case, ignoring the cosmic variance contribution to the covariance, we can analytically estimate the ratio of the Fisher matrix $F^{(T)}_{EE}/F^{(T)}_{BB}$ by
\begin{align}
\frac{F^{(T)}_{EE}}{F^{(T)}_{BB}} &
\approx
\frac{\int^{1}_{-1}{\rm d}\mu\, P^{(T)}_{EE}}
{\int^{1}_{-1}{\rm d}\mu\, P^{(T)}_{BB}}
\notag \\
&
\approx
\frac{
\left(\frac{1}{16}\int^{1}_{-1}{\rm d}\mu\, (1+\mu^{2})^{2}\right)^{2}}
{\left(\frac{1}{4}\int^{1}_{-1}{\rm d}\mu\, 
\mu^{2}\right)^{2}}
\notag \\
& = \frac{83}{63} ,
\end{align}
yielding $\sigma_{EE}(r_{T}) \approx \sqrt{63/83} \, \sigma_{BB}(r_{T}) = 0.87 \, \sigma_{BB}(r_{T})$. Similarly, in the vorticity vector mode case, we have $\sigma_{EE}(r_{V}) \approx \sqrt{21}\, \sigma_{BB}(r_{V}) = 4.6 \, \sigma_{BB}(r_{V})$. These values explain the results in Table~\ref{tab: Euclid / SKA vector and tensor} very well.
We note that the minimum detectable value given in Table~\ref{tab: Euclid / SKA vector and tensor} is larger than the constraints derived in Ref.~\cite{2023arXiv230908653P} because our analysis assumes the spectroscopic survey which has generally smaller number of galaxies than the photometric surveys assumed in Ref.~\cite{2023arXiv230908653P}.

In Fig.~\ref{fig: primordial V and T}, we perform the same analysis as in Fig.~\ref{fig: Bl mag ideal} but for the cases of the vorticity vector mode and primordial GWs.
In analogy with the magnetic case in Fig.~\ref{fig: Bl mag ideal}, as the shot noise decreases, the detectability of $r_{V/T}$ from the E- and B-modes reaches a plateau because of the scalar-mode contamination in the covariance. In a noisy regime as $10^{0} \, ({\rm Mpc}/h)^{3} \lesssim \sigma^{2}_{\gamma}/n_{\rm gal}$, as the above analytic estimate indicates, the E-mode spectrum can capture smaller $r_T$ than the B-mode one.

\bibliographystyle{apsrev4-2}
\bibliography{ref}
\end{document}